\documentclass{aa}

\usepackage{graphicx}
\usepackage[varg]{txfonts}
\usepackage{xcolor}
\usepackage{arydshln}

\bibpunct{(}{)}{;}{a}{}{,} 

\usepackage{amssymb}
\usepackage{natbib}
\begin{document}

\title{$^{44}\rm Ti$ ejecta in young supernova remnants}

\author{Christoph Weinberger \inst{1} \and Roland Diehl \inst{1} \and Moritz M. M. Pleintinger \inst{1} \and Thomas Siegert \inst{2} \and Jochen Greiner \inst{1}}

\institute{Max-Planck-Institut für extraterrestrische Physik, Gießenbachstraße, 85741 Garching, Germany \\ e-mail: cweinb@mpe.mpg.de \and
Center for Astrophysics and Space Sciences, University of California, San Diego, 9500 Gilman Dr, La Jolla, CA 92093, USA}

\date{Received DD Month YYYY;\
 accepted DD Month YYYY}

\abstract
{Tracing unstable isotopes produced in supernova nucleosynthesis provides a direct diagnostic of supernova explosion physics. Theoretical models predict an extensive variety of scenarios, which can be constrained through observations of the abundant isotopes $^{56}\rm Ni$ and $^{44}\rm Ti$. Direct evidence of the latter was previously found only in two core-collapse supernova events, and appears to be absent in thermonuclear supernovae.}
{We aim to to constrain the supernova progenitor types of Cassiopeia A, SN 1987A, Vela Jr., G1.9+0.3, SN1572, and SN1604 through their $^{44}\rm Ti$ ejecta masses and explosion kinematics.}
{We analyzed INTEGRAL/SPI observations of the candidate sources utilizing an empirically motivated high-precision background model. We analyzed the three dominant spectroscopically resolved de-excitation lines at 68, 78, and 1157\,keV emitted in the decay chain of $^{44}\rm Ti$  $\rightarrow$ $^{44}\rm Sc$ $\rightarrow$ $^{44}\rm Ca$. The fluxes allow the determination of the production yields of $^{44}\rm Ti$. Remnant kinematics were obtained from the Doppler characteristics of the lines.}
{We find a significant signal for Cassiopeia A in all three lines with a combined significance of 5.4$\sigma$. The fluxes are $(3.3 \pm 0.9) \times 10^{-5}\,\mathrm{ph\,cm^{-2}\,s^{-1}}$, and $(4.2 \pm 1.0) \times 10^{-5}\,\mathrm{ph\,cm^{-2}\,s^{-1}}$ for the $^{44}\rm Ti$ and $^{44}\rm Sc$ decay, respectively. This corresponds to a mass of $(2.4 \pm 0.7)  \times 10^{-4}\,\mathrm{M_{\odot}}$ and $(3.1 \pm 0.8) \times 10^{-4}\,\mathrm{M_{\odot}}$, respectively. We obtain higher fluxes for $^{44}\rm Ti$ with our analysis
of Cassiopeia A than were obtained in previous analyses. We discuss potential differences. We interpret the line width from Doppler broadening as expansion velocity of $(6400 \pm 1900)\,\mathrm{km\,s^{-1}}$. We do not find any significant signal for any other candidate sources.}
{We obtain a high $^{44}\rm Ti$ ejecta mass for Cassiopeia A that is in disagreement with ejecta yields from symmetric 2D models. Upper limits for the other core-collapse supernovae are in agreement with model predictions and previous studies. The upper limits we find for the three thermonuclear supernovae (G1.9+0.3, SN1572 and SN1604) consistently exclude the double detonation and pure helium deflagration models as progenitors.}


\maketitle

\section{Introduction}\label{sec:Introduction}

Supernova explosions play a crucial role in the chemical and kinematic evolution of the Universe. Self-consistent detailed models for the explosion mechanism and the ensuing kinematics of the ejected material are still lacking. Despite the frequent occurrence of supernova explosions ($1.9 \pm 1.1$ per century in the Galaxy \citep{Diehl2006}), observational constraints on the explosion mechanisms are still sparse, due to the large variety of progenitor systems and large parameter space of their models.
\\
For modeling core-collapse supernova explosions, reviving the stalled shock and triggering an explosion presents a major challenge \citep{Janka2012,Burrows2018}, as it has long been understood that the prompt explosion mechanism following core bounce cannot explode the star. Energy deposition by neutrinos in a gain region close to the stalled shock is considered as the driving force of the shock revival. Heating-induced creation of hydrodynamic effects, observed in 2D and 3D models, such as neutrino-driven convection and the standing accretion shock instability (SASI) \citep{Blondin2007,Marek2009,Bruenn2013,Hix2016,Bruenn2016} enhance the explodability, however, different implementation schemes favor the dominance of either neutrino-driven convection or the SASI \citep{Pan2016,Summa2016,Skinner2016}. Inclusion of additional, microphysical effects, for example, strangeness corrections \citep{Melson2015} and rotation from the progenitor \citep{Mueller2017,Summa2018,Takiwaki2016,Iwakami2014} affect the explodability. Model calculations of supernova explosions performed by various groups lead to successful explosions in a mass range $8 \lesssim M/\mathrm{M_{\odot} \lesssim 25}$. Asymmetries evolved from hydrodynamic effects become frozen in the explosion and are reflected in the kinematics and mass distribution of the ejecta \citep{Nomoto1995,Buras2006,Fryer2006,Takiwaki2012,Wongwathanarat2015,Orlando2016,Wongwathanarat2017}.
\\
The variety of observed type Ia supernova luminosity and temporal behavior of light curves in the first few hundred days \citep{Phillips1999} cannot be reproduced by a single progenitor type \citep{Wang2012,Hillebrandt2013}. Mergers of binary white dwarfs \citep{Kerkwijk2010,Pakmor2010,Ruiter2012,Pakmor2013,Kashyap2018} and mass accretion on single white dwarf stars (single degenerate scenario) are expected to lead to a central thermonuclear runaway, disrupting the white dwarf star in the process. On single degenerates, both stable mass accretion on white dwarfs towards the Chandresekhar mass limit \citep{Nomoto1984,Parthasarathy2007,Maeda2010,Woosley2011,Hachisu2011,Seitenzahl2013,Chen2014,Fink2014} and surface helium detonation \citep{Livne1990,Fink2010,Shen2014,Leung2020} can increase the central density sufficiently to ignite nuclear fusion.
\\
The high opacity of the ejected material does not allow direct observations of the first stages of the explosions. One of the most promising methods of deducing physical constraints for the explosion mechanisms is to directly observe the decay of radioactive isotopes produced during explosive nucleosynthesis. The best candidates for observing supernova interior physics through their nucleosynthesis imprints are the isotopes $^{56}\rm Ni$ and $^{44}\rm Ti$ due to their high abundances ($\approx 10^{-2}$ $\mathrm{M_{\odot}}$ for core-collapse supernovae, 0.5 $\mathrm{M_{\odot}}$ for SNe Ia in $^{56}\rm Ni$, $10^{-6} -10^{-4}$$\mathrm{M_{\odot}}$ in $^{44}\rm Ti$, see also below). Both isotopes have, for gamma-ray observations, ideal radioactive decay times: 6\,d and 77\,d for $^{56}\rm Ni$ and its daughter nucleus $^{56}\rm Co$, respectively, allow only a very short post-explosion observation window in which the majority of emitted gamma-rays are expected to be absorbed by the dense stellar material. With a half life of $58.9\pm0.3$\,yr \citep{Ahmad2006}, a considerable amount of $^{44}\rm Ti$ is still decaying even after centuries when the supernova remnant has long become optically thin to X- and gamma-ray emission. In contrast to the decay of synthesized $^{26}\rm Al$ and $^{60}\rm Fe$ with half lives of 700\,kyr and 2\,Myr, producing a diffuse emission throughout the Milky Way \citep{Plueschke2001,Bouchet2015,Siegert2017}, the emission signature of $^{44}\rm Ti$ decay is expected to be the one of individual point sources. 
\\
Spatially co-produced in core-collapse supernovae, $^{44}\rm Ti$ and $^{56}\rm Ni$ are mainly synthesized during alpha-rich freeze out \citep{Woosley1973} deep in the central region of the supernova, where nucleosynthesis is strongly dependent on the thermodynamic conditions of the inner ejecta \citep{Magkotsios2010}. Due to the hydrodynamic instabilities required for an effective explosion, mixing and asymmetric expansion of the burning volume invalidates the 1D model characteristics of a mass cut, an idealized radius \citep{Woosley1995}, separating the gravitationally bound material from the ejecta. Since the nuclear burning occurs close to such a mass cut, the final amount of ejected $^{44}\rm Ti$ is subject to uncertainty and ranges between $10^{-5}-10^{-4}$ M$_{\odot}$ \citep{Timmes1996,Limongi2018}.
\\
Type Ia supernovae typically produce $\approx 10^{-6}$ M$_{\odot}$ $^{44}\rm Ti$ in centrally ignited pure deflagration models, and up to a few times $10^{-5}$ M$_{\odot}$ in delayed detonation models \citep{Maeda2010,Seitenzahl2013,Fink2014}. Double detonation and surface He deflagration of sub-Chandrasekhar mass white dwarfs produce yields up to $10^{-3}\rm\,M_{\odot}$ of $^{44}\rm Ti$ \citep{Fink2010,Woosley2011,Moll2013}. The subclass of helium surface explosions synthesizes $^{56}\rm Ni$ very little, potentially producing subluminous type Ia supernova events. Simulations \citep{Waldman2011} and observations \citep{Perets2010} suggest special configurations of binary white dwarfs, which, when exploding as peculiar type Ia supernovae, can produce of the order of $10^{-2}\rm\,M_{\odot}$ of $^{44}\rm Ti$. 
\\
Observations of late optical spectra allow for the determination of mass ratios of synthesized elements, constraining the burning mechanisms \citep{Eriksen2009,Pakmor2010,Jerkstrand2015,Maguire2018,Mori2018}. However, the total ejected mass remains a free parameter. Nucleosynthesis yields can be estimated from the bolometric light curves of explosion, however, inferring the ejected mass of radioactive material is highly model dependent \citep{Seitenzahl2014}.
 Comparing optical and infrared spectra to the late-time light curve, an ejected $^{44}\rm Ti$ mass of $(1.5 \pm 0.5)  \times 10^{-4}\,\mathrm{M_{\odot}}$ is obtained for SN 1987A \citep{Jerkstrand2015} in agreement with NuSTAR findings of $(1.5 \pm 0.3)  \times 10^{-4}\,\mathrm{M_{\odot}}$ \citep{Boggs2015} based on the 68 and 78\,keV line. This is, however, in disagreement with the results obtained from a multicomponent long-time light-curve model of $(0.6 \pm 0.2)  \times 10^{-4}\,\mathrm{M_{\odot}}$ \citep{Seitenzahl2013} and direct detection of $^{44}\rm Ti$ decay with INTEGRAL/IBIS at 68 and 78\,keV of $(3.1 \pm 0.8)  \times 10^{-4}\,\mathrm{M_{\odot}}$ \citep{Grebenev2012} .
\\
Direct observational evidence for the production of $^{44}\rm Ti$ can be obtained through the decay chain of $^{44}\rm Ti$ $\rightarrow$ $^{44}\rm Sc$ $\rightarrow$ $^{44}\rm Ca,$ where the prominent decay lines arise at energies of 68\,keV and 78\,keV for the decay of $^{44}\rm Ti,$ and at 1157\,keV for the $^{44}\rm Sc$ decay. These lines have a probability of 93.0\,$\%$, 96.4\,$\%$, and 99.9\,$\%$ per decay, respectively \citep{Chen2011}. In addition, a fluorescence photon is emitted from shell transitions in $^{44}\rm Sc$ with a probability of 16.7\,$\%$ at 4.1\,keV. With a significant difference of half life times of 60\,yr in the first decay and 4\,h \citep{Audi2003} in the subsequent decay, the activity of all three decay channels can be safely assumed to be identical after correcting for the branching ratios.
In this work, we utilized INTEGRAL/SPI data to search for emission of $^{44}\rm Ti$ in all three decay lines simultaneously in the young close by supernova remnants Cassiopeia A, Tycho, Kepler, G1.9+0.3, Vela Junior, and the extragalactic (but very young) SN 1987A. We aim to constrain both $^{44}\rm Ti$ ejecta yields and explosion kinematics  for remnants with ages of less than a few centuries where a substantial amount of $^{44}\rm Ti$ may still be present. We re-evaluated previous analyses of these remnants concerning the decay of $^{44}\rm Ti$ (see Section \ref{sec:remnants}) in the 68 and 78\,keV lines and improved on them by including decay signature of the $^{44}\rm Sc$ daughter nucleus at 1157\,keV, which can only be seen in SPI. The 1157\,keV line was studied for the remnant Cassiopeia A and Vela Jr.
\\
This paper is structured as follows: Section \ref{sec:remnants} provides an overview of the six target remnants. Section \ref{sec:Data_Analysis} describes SPI data and spectral analysis, followed by Section \ref{sec:results}, which includes results for the six remnants. Finally, in Section \ref{sec:discussion}, we discuss our results and give a summary of astrophysical implications.

\section{Young supernova remnants}\label{sec:remnants}

\begin{table*}\centering

\caption{Astrophysical parameters for the analyzed objects including the distances of supernova remnants used in the mass estimate for $^{44}\rm Ti$. The age of the remnant is given for 01.01.2011\,AD, which is the average observation date also used in Eq. \ref{eq:decaylaw}. Age uncertainties of the time bin are included in the mass estimates. Distance to Vela Jr. is estimated from $^{44}\rm Ti$ yields. Distance to G1.9+0.3 is estimated from absorption towards the Galactic center. Exposure is dead time corrected, effective INTEGRAL/SPI exposure. References: (1) \citep{Alarie2014}; (2) \citep{Reed1995}; (3) \citep{Krause2008}; (4) \citep{Pietrzynski2019}; (5) \citep{Gall2015}; (6) \citep{Aschenbach1999}; (7) \citep{Hayato2010}; (8) \citep{Krause2008a}; (9) \citep{Sankrit2016}; (10) \citep{Reynolds2007}; (11) \citep{Reynolds2008}; (12) \citep{Borkowski2013}.}
\label{table:SN_remnant}
\begin{tabular}{c c c c c c c}
\hline\hline
                                                                & Cassiopeia A               & SN 1987A               & Vela Jr                       & Tycho                   & Kepler                                        & G1.9+0.3                \\ 
\hline 
Distance $[\mathrm{kpc}]$               		& $3.3\pm0.1$   			& $49.6\pm0.5$  &$ 0.2$                    & $4.1\pm1$             & $5.1^{+0.8}_{-0.7}$           & $8.5$                   \\ 
Year of Explosion                                       	& 1681                          	& 1987                    & ~1320                         & 1572                          & 1604                                            & $1890$                \\
Age $[\mathrm{yr}]$                             	& 330                           & 24                              & 690                           & 438                           & 406                                             & 120                   \\
Type                                                    	& IIb                           & II-P                            & II                            & Ia                                    & Ia                                              & Ia                            \\ 
Exposure $[\mathrm{Ms}]$               		& 11.2                          & 7.0                             & 8.3                           & 10.3                          & 29.3                                            & 30.6                  \\
Galactic Coordinates; l/b [deg] 				& $111.7/-2.1$          	& $279.7/-31.2$ & $266.3/-1.21$   & $120.1/1.4$           & $4.5/6.8$                             & $1.9/0.3$                       \\
References                                                 	& 1, 2, 3                       & 4, 5                            & 6                                     & 7, 8                            & 9, 10                                         & 11, 12                  \\
\hline
\end{tabular}

\end{table*}

The astrophysical parameters of the six most promising candidates to coherently observe the $^{44}\rm Ti$ decay chain are listed in Tab. \ref{table:SN_remnant}. Evidence for the signature of the $^{44}\rm Ti$ decay has been claimed in the majority of these historic supernova explosions. Within SPI's narrow-line sensitivity verification of the $^{44}\rm Ti$ signal and detection of the ensuing decay of $^{44}\rm Sc$ is feasible within a limited age-distance volume including these remnants. We searched for signatures of the $^{44}\rm Ti$ decay at 68 and 78\,keV due to the de-excitation of $^{44}\rm Sc$*, and at 1157\,keV due to the subsequent de-excitation of $^{44}\rm Ca$*. We utilized data from the INTEGRAL mission from 2003 to 2019, meaning INTEGRAL revolutions 43 to 2047. For our study, we applied an average date of 01.01.2011\,AD as an observation date for all calculations and included age uncertainties in our mass derivation.

\subsection{Cassiopeia A}

Cassiopeia A is one of the best studied supernova remnants in the Milky Way \citep{Vink2004}. With an approximate age of 340 years, it is the youngest known Galactic supernova remnant, attributed to a core-collapse explosion. From the detection of hydrogen and weak helium lines in a supernova light echo, attributed to Cassiopeia A, the explosion is characterized as a type IIb supernova \citep{Krause2008}. Explosions of this type are typically produced by 15--25\,$\mathrm{M_{\odot}}$ stars \citep{Young2006}. The decay of $^{44}\rm Ti$ has been consistently measured \citep{Siegert2015,Grefenstette2014,Iyudin1994,Vink2001} from Cassiopeia A with an average inferred $^{44}\rm Ti$ ejecta mass of $(1.37 \pm 0.19)  \times 10^{-4}\,\mathrm{M_{\odot}}$.

\subsection{SN 1987A}

The explosion of SN 1987A occurred in the Large Magellanic Cloud (LMC) on February 24 1987, giving rise to a peculiar light curve containing a plateau phase (type II-P supernova). Detection of $^{56}\rm Co$ lines \citep{Matz1988,Tueller1990} for the first time directly has confirmed that supernova light is indeed powered by this isotope, produced in the inner regions at the time of core-collapse. Recent refinements of astronomical precision yield a distance of $49.6\pm0.5$\,kpc \citep{Pietrzynski2019} to the LMC. Direct proof for the decay of $^{44}\rm Ti$ has been reported from measurements with IBIS/INTEGRAL and NuSTAR in the hard X-ray lines of $^{44}\rm Sc$. However, the two measurements show a discrepancy in the ejecta mass, with $(3.1 \pm 0.8)  \times 10^{-4}\,\mathrm{M_{\odot}}$ and $(1.5 \pm 0.3)  \times 10^{-4}\,\mathrm{M_{\odot}}$, respectively \citep{Grebenev2012,Boggs2015}. NuSTAR constrains the ejecta kinematics to an expansion velocity of less than 4100\,$\rm km\,s^{-1}$, and a bulk Doppler shift suggests an asymmetric explosion.

\subsection{Vela Jr.}

COMPTEL discovered a significant gamma-ray emission in the energy range centered at the 1157\,keV $^{44}\rm Ca$ line. The signal was located in the direction of the Vela region with a flux of $(3.8 \pm 0.7) \times 10^{-5}\,\mathrm{ph\,cm^{-2}\,s^{-1}}$ \citep{Iyudin1998}. This emission was attributed to the decay of $^{44}\rm Ti$ in a previously unknown type II supernova remnant, which had been identified through detailed analysis of the X-ray emission of this region, and was called RX J0852.0-4622 or "Vela Jr." \citep{Aschenbach1998}. An estimated age of $\approx$680\,yr and a distance of $\approx$200\,pc is derived \citep{Aschenbach1999}. The remnant has an apparent diameter of 2$^{\circ}$ \citep{Aharonian2007}. Follow-up observation with ASCA and XMM-Newton report a line at 4.4keV \citep[][respectively]{Tsunemi2000,Iyudin2005} possibly from $^{44}\rm Sc$ fluorescence. However, the more likely remnant age of 2.4-5.1\,kyr derived from the expansion rate of the supernova \citep{Allen2014} would exclude detectability of $^{44}\rm Ti$ decay emission. This would then also be in agreement with the upper limits determined with IBIS/INTEGRAL \citep{Tsygankov2016}.

\subsection{G1.9+0.3}

G1.9+0.3 is presumably the youngest supernova remnant seen in the Galaxy so far. It has been identified by \citet{Reynolds2008} in the years 1985 and 2008 in the radio and X-ray regime. It is identified as a type Ia explosion. Radio observations suggest a distance of 8.5\,kpc, placing the remnant in the Galactic center \citep{Reynolds2008}. From the apparent increase in size, an age of $\approx 100$\,yr is deduced, which suggests a very high expansion velocity of 14000\,km\,s$^{-1}$ for the shock front. Using Chandra data, a detection of a soft X-ray component at 4.1\,keV from the fluorescence line of $^{44}\rm Sc$ has been reported \citep{Borkowski2010} inferring a $^{44}\rm Ti$ ejecta mass of $(1-7) \times 10^{-5}$\,$\mathrm{M_{\odot}}$. Extrapolating the detected flux in the $^{44}\rm Sc$ fluorescence line to the hard X-ray lines provides a line flux estimate for the 68\,keV $^{44}\rm Ti$ line, which is below the upper limits determined from NuSTAR and IBIS/INTEGRAL instruments \citep[$(0.7-1.5)\times 10^{-5}$\,ph\,cm$^{-2}$\,s$^{-1}$, $9\times 10^{-6}$\,ph\,cm$^{-2}$\,s$^{-1}$; ][respectively]{Zoglauer2015,Tsygankov2016}.

\subsection{Tycho}

The Tycho supernova remnant is attributed to an explosion of type Ia, from measurements of the light echo and a comparison of model light curves with X-ray spectra \citep{Badenes2006,Krause2008}. Tycho exploded in 1572 AD. Detection of synchrotron emission from a thin shell supports the idea of particle acceleration in young supernova remnants \citep[e.g.,][]{Slane2014}. X-ray line emission from intermediate to iron group elements in the interior of the supernova remnant has been found to be clumped \citep[XMM Newton][]{Miceli2015}, and the hard X-ray lines of the $^{44}\rm Ti$ decay chain were also detected \citep[Swift/BAT][]{Troja2014}. These measurements suggest the presence of titanium both in the shocked shell and the interior region of the remnants. However, upper limits obtained from NuSTAR measurements exclude the presence of $^{44}\rm Ti$ within a 2' remnant radius at the Swift/BAT detection level over a large range of expansion velocities \citep{Lopez2015}. The distance to the remnant is somewhat uncertain, with estimates ranging from 1.7 to 5.1\,kpc \citep{Hayato2010,Slane2014,Albinson1986,Voelk2008}. We adopted a distance of $4.1\pm1$\,kpc \citep{Hayato2010}.

\subsection{Kepler}

Johannes Kepler detected this supernova in 1604\,AD. Also named G4.5+6.8, this is the youngest Galactic supernova for which an optical transient has been observed. This supernova occurred at a distance of $5.1\pm0.8$\,kpc \citep{Sankrit2016}, and is located $460-600$\,pc above the Galactic plane. Due to the detection of strong iron lines in the ejecta, it is attributed to a type Ia explosion \citep{Reynolds2007}, also supported by the absence of a central compact object. $^{44}\rm Ti$ decay lines have not been found, with upper limits of $1.8 \times 10^{-5}$\,ph\,cm$^{-2}$\,s$^{-1}$ in the 1157\,keV line \citep{Dupraz1997} determined from COMPTEL data and $6.3 \times 10^{-6}$\,ph\,cm$^{-2}$\,s$^{-1}$ determined from INTEGRAL/IBIS for the 68 and 78\,keV line \citep{Tsygankov2016}.

\section{Data and analysis method}\label{sec:Data_Analysis}

\subsection{Instrument and analysis method}

\begin{figure}
        \includegraphics[width=\linewidth]{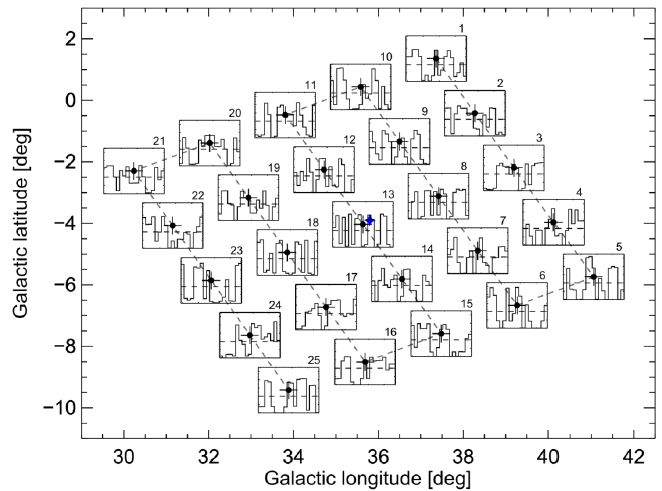}
        \caption{Intensity distribution of a celestial source located at the center of the plot (blue dot) \citep{Siegert2019}. The orientation of the SPI instrument is changed in steps of two degrees to visit the locations marked on the map with black dots. The instrument remains centered on each celestial position for $\approx$2000s, which is called "one pointing". The source located at the blue position is folded through the instrumental response function to determine the intensity distribution in the detector plane.}
  \label{fig:Patterns}
\end{figure}

ESA's gamma-ray space observatory INTEGRAL \citep{Winkler2003} carries two main instruments on board, the imager IBIS and the spectrometer SPI. The SPI camera \citep{Vedrenne2003} is a germanium detector array consisting of 19 hexagonally shaped detectors, optimized for high-resolution spectroscopy in the energy range between 18\,keV and 8\,MeV, with a spectral resolution of $\approx 2.3\,\mathrm{keV}$ (full width at half maximum, FWHM) at 1\,MeV. SPI electronics records 16384 energy channels in the range of $18\,\mathrm{keV} - 2\,\mathrm{MeV}$, which is called SPI's "low-energy range". We analyzed SPI data for the signatures of the $^{44}\rm Ti$ decay lines, which have centroid energies in the laboratory at 68, 78\,keV and 1157\,keV. Our analysis is focused on the energy bands $30-100\,\mathrm{keV}$ and $1090-1210\,\mathrm{keV}$ in order to constrain potential underlying continuum emission at lower energies and account for potential large line broadening at higher energies.
\\
SPI data after initial energy calibration and pre-processing comprises spectra in 0.5\,keV bins for each of the 19 detectors, accumulated over exposures of typically 2000\,s, called pointings. The orientation of the satellite is shifted by $\approx 2^{\circ}$ between each consecutive exposure in a rectangular-shaped dithering pattern consisting of 5$\times$5 sets of coordinates around the observation target. We include data in our analysis in which the celestial objects of interest are within the partially coded field of view of SPI of $34^{\circ}\times34^{\circ}$. In general, celestial photons entering the aperture of SPI are partially blocked by a coded mask placed 171\,cm above the detector plane, imprinting shadow grams on the camera. For an idealized source at long integration times, this creates relative detector intensity distributions (detector patterns), since the absolute number of measured photons per detector is governed by the visibility of the source through the mask (Fig. \ref{fig:Patterns}). Sources within a field of view of $16^{\circ}\times16^{\circ}$ are fully coded by the tungsten mask, with decreasing coding fraction towards the coding limit at $34^{\circ}\times34^{\circ}$.
\\
The main challenge of SPI data analysis consists of extracting the sparsely populated celestial detector pattern above a large, underlying, instrumental background. The latter is introduced by interaction and activation of satellite and instrument material by cosmic ray bombardment. In our spectroscopic analysis method, we compare the combination of the celestial detector patterns and the background detector patterns to the time series of measured patterns for the 19 detectors by fitting (time dependent) scaling parameters for both contributors. The celestial detector intensity distribution is calculated by applying the energy-dependent image response function (IRF) specific to SPI's tungsten mask and the re-orientations during the dithering exposures. The model of the background consists of two separate components, one for continuum emission and one for nuclear de-excitation lines at specific energies. Both components are determined over a broad energy range and from multiple years of data. Degradation effects of detectors and time dependent variance of background level are taken into account by modeling the background per detector on an orbital timescale. 
The mathematical description of our modeling method is given by
\begin{equation} \label{eq:m_k}
        m_k = \sum_j R_{j,k} \sum_{i}^{N_{l}} \theta_i M_{i,j} +  \sum_{i}^{N_{b}} \theta_{i} B_{j,k}      
,\end{equation}
which means that the data and model per energy bin \textit{k} are represented by the sum of the celestial components \textit{i} of the total number of $N_{l}$ celestial sources convolved through the image response function \textit{R} per detector \textit{j} and the sum of all background components $N_{b}$ of detector \textit{j}. No prior knowledge concerning the energy spectrum of the celestial sources is assumed in our analysis. In general, the model is fit to the data by minimizing the Cash Statistic \citep{Cash1979}, adjusting the scaling parameters $\theta_i$ in Eq. (\ref{eq:m_k}), where different timescales for the scaling of the components are allowed. We use the spimodfit analysis tool \citep{Strong2005,Halloin2009}, which applies a Levenberg-Marquardt algorithm to determine the maximum likelihood solution for all intensity parameters $\theta_i$. The software is based on the ISDC software spiros \citep{Dubath2005}, however optimized for high spectroscopic resolution of low signal to noise sources. Unless otherwise stated, uncertainties are given as $1\sigma$. We use the Pearson $\chi^2$ as an absolute goodness-of-fit criterion. We note that the chosen absolute goodness-of-fit criterion ($\chi^2 \stackrel{!}{=} 1.0$) itself carries an uncertainty \citep{Andrae2010}.The smallest possible timescale we utilized is a single pointing. To minimize the contamination of our data set by known periods of increased background, we excluded orbit phases below 0.10 and above 0.88, during which the satellite passes through the Van Allen radiation belts.

\subsection{Background modeling} \label{sec:bgmodeling}

SPI instrumental background mainly originates from the bombardment of the satellite by cosmic ray particles. Interaction of these cosmic rays can induce nuclear reactions in the satellite materials. Subsequent decays from excited nuclear levels lead to the emission of nuclear de-excitation lines, which fall into the energy range of the SPI detectors. Among others, bremsstrahlung is a second dominant contribution to the background, forming an underlying continuum.
\\
To determine the temporal and spectral behavior of the background, we used the knowledge gained from 17\,yr of integrated mission data. Long-term temporal variation is introduced by the degradation of the lattice structure of the Germanium detectors and the absolute production rate of cosmic rays, which is anticorrelated with the solar cycle. Short-term variations are introduced by solar flares.\\

The detailed spectral shapes of the continuum emission and nuclear de-excitation lines are determined separately. Since a physically based model is difficult to construct and calibrate at the required precision, we used an empirical description of the background. This is based on previous attempts to model the highly variable instrumental background in SPI, as, for example, in \citet{Knoedlseder2004} and\citet{Jean2003}, and supersedes the standard on-off methods as presented in \citet{Dubath2005}. All background components are determined as a linear superposition of an underlying continuum normalized to a central pivot energy, superimposed by emission lines. The line shapes are represented by Gaussian functions convolved with a degradation function, which accounts for the degradation of the germanium charge collection efficiency \citep{Kretschmer2011}.
We determine the spectral shape on a 3\,d (one orbit) period separately for each detector to trace the time-dependent degradation of the detectors. This timescale is chosen as the best compromise between accumulating sufficient statistics and appropriate determination of temporal variations of the spectral shape. Secondary contributions to the background are smeared out in our coded-mask analysis by accumulating data over multiple pointings \citep{Siegert2019}. The consistency of this high-resolution, time-dependent background modeling approach is demonstrated, for example, in \citet{Siegert2016,Siegert2017b}, and \citet{Diehl2018}.

\subsection{Spectral analysis}

With the modest spatial resolution of SPI of $\approx3^{\circ}$, supernova remnants cannot be resolved in separate clumps of ejecta. To enhance sensitivity for the relatively low-intensity total celestial signal in a line, a model for the line shape has to be adopted. We describe the emission produced by radioactive decay with Gaussian shaped lines, plus a power-law-shaped continuum. 
\begin{equation}
        LS(E;E_0,F_0,\sigma) = \frac{F_0}{\sqrt{2\pi}\sigma}\cdot\mathrm{exp}\left(\frac{(E-E_0)^2}{2\sigma^2}\right) + A_0\cdot \left(\frac{E}{E_C}\right)^{\alpha}
,\end{equation}
where $F_0$ is the measured line flux, $E_0$ is the energy of the Doppler shifted-line centroid, and $\sigma$ is line width. We interpret any broadening of the line, which would be additional to the detector resolution, as Doppler broadening caused by the expansion velocity of the ejecta. We determine the line parameters separately for each line when possible. No changes in the kinematics of the ejecta are expected, allowing for a combined three-line fit, assuming identical Doppler parameters for all three lines. We further allow for an underlying celestial continuum accounting for bremsstrahlung processes, with normalization parameter $A_0$ and power-law index $\alpha$. The mass of the ejected $^{44}\rm Ti$ per line is determined by
\begin{equation} \label{eq:decaylaw}
M_{Ti_{44}} = F_{L}\cdot 4\pi d^2\cdot N\cdot u\cdot \tau\cdot\mathrm{exp}(t/\tau)
,\end{equation}
where $F_{L}$ is the flux of the specific line $L = (68,78,1157)$\,keV, $d$ is the distance to the source, $N = 44$ is the number of nuclei in $^{44}\rm Ti$, $u$ is the atomic mass number, $\tau = 86.6$\,yr is the decay constant of $^{44}\rm Ti,$ and $t$ is the age of the supernova remnant. All line fluxes $F_{0}$ are normalized with the branching ratio $b_L$\footnote{This is equal to the probability of photon emission per decay (93.0\,\%, 96.4\,\%, 99.9\,\%, respectively; s. Sect. \ref{sec:Introduction})} of the specific line $L$. For comparison, all flux values are stated as the normalized flux $F_L = F_0/b_L$. In a multiline fit, the branching ratios determine the relative line intensities, and the normalized flux $F_L$ is fit as the parameter of interest. Velocities corresponding to the Doppler broadening of the lines are calculated for the ejecta from the FWHM of the line, assuming a uniformly expanding sphere. While this model might not adequately describe asymmetries as seen in the supernova remnant Cassiopeia A \citep{Grefenstette2017}, it provides a reasonable first-order approach for determining fluxes from the remaining unresolved sources. Confidence intervals for our results of the spectral fits are estimated from the 68th percentile interval of a Metropolis-Hastings algorithm, minimizing the Pearson $\chi^2$ as test statistics. Upper limits are given at $2\sigma$ (i.e., $\Delta\chi^2 = 4$, for one degree of freedom, dof). We derived our upper limits by varying only the integrated flux of the respective line, assuming values for Doppler broadening and shift, and utilizing the best fit values for the underlying continuum.

\section{SPI results}\label{sec:results}

\subsection{Cassiopeia A} \label{Sec:CasA}

\begin{figure}
        \includegraphics[width=\linewidth,trim={1.5cm 1.5cm 1.5cm 1.5cm}]{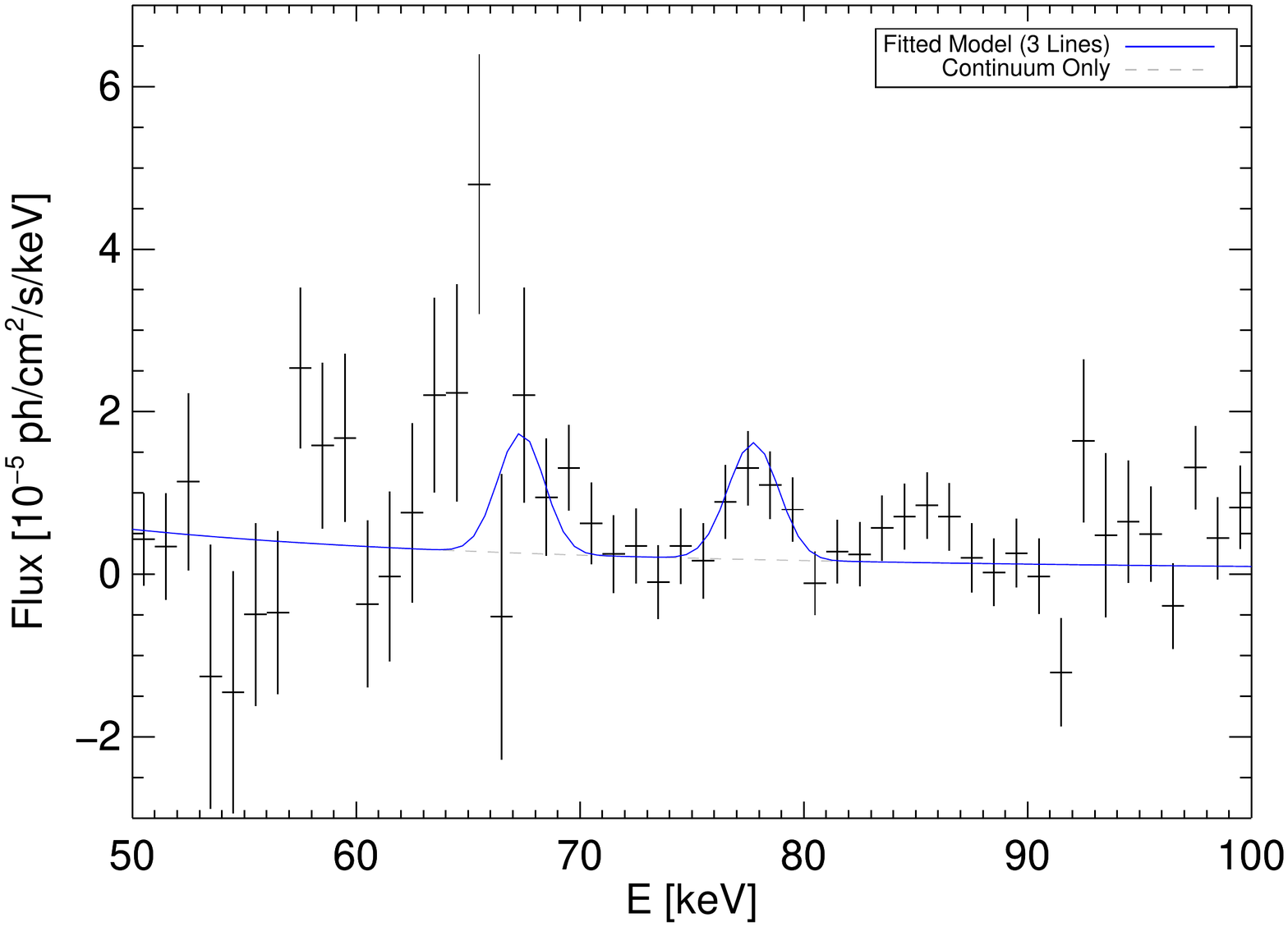}
        \includegraphics[width=\linewidth,trim={1.5cm 1.5cm 1.5cm 1.5cm}]{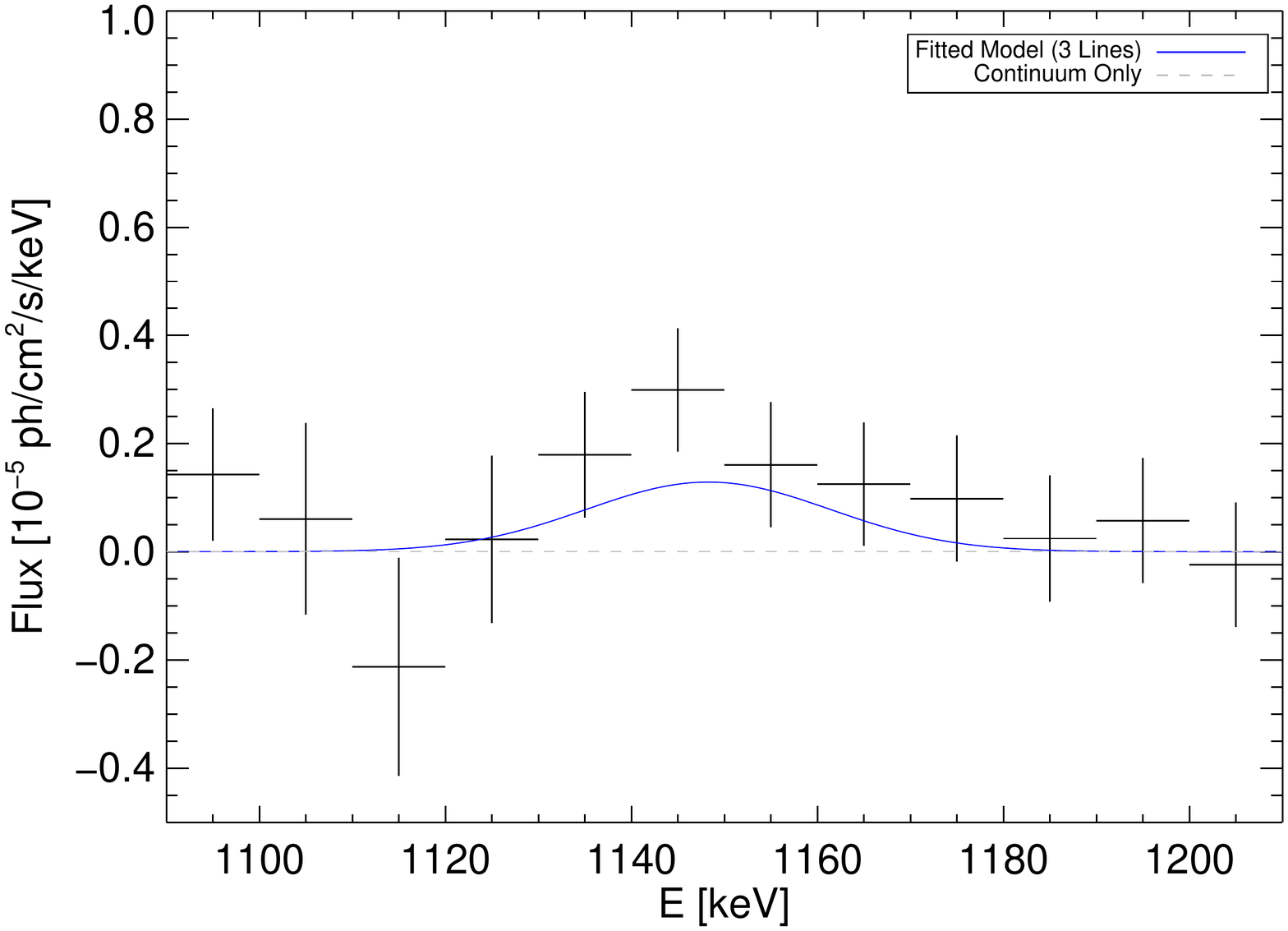}
        \caption{Spectra for Cassiopeia A in the energy region 50 -- 100\,keV in 1\,keV binning and 1090 -- 1210\,keV in 10\,keV binning containing all potential decay lines from the decay chain of $^{44}\rm Ti$. We fit the spectrum with a power law accounting for underlying continuum most likely produced by synchrotron emission at the shock front and Gaussian shaped line profiles in each region. The lines are Doppler shifted and broadened, which cannot solely be explained by the instrumental resolution at the energies, respectively. The lines are determined from a combined fit. This means that we fit one uniform power law over the entire energy range and identical Doppler parameters and integrated flux values for all three lines simultaneously. Flux values are corrected for the respective branching ratios of the lines.} 
  \label{CasALELines}
\end{figure}

\begin{figure}
        \includegraphics[width=\linewidth,trim={2.5cm 1.5cm 1.5cm 1.5cm}]{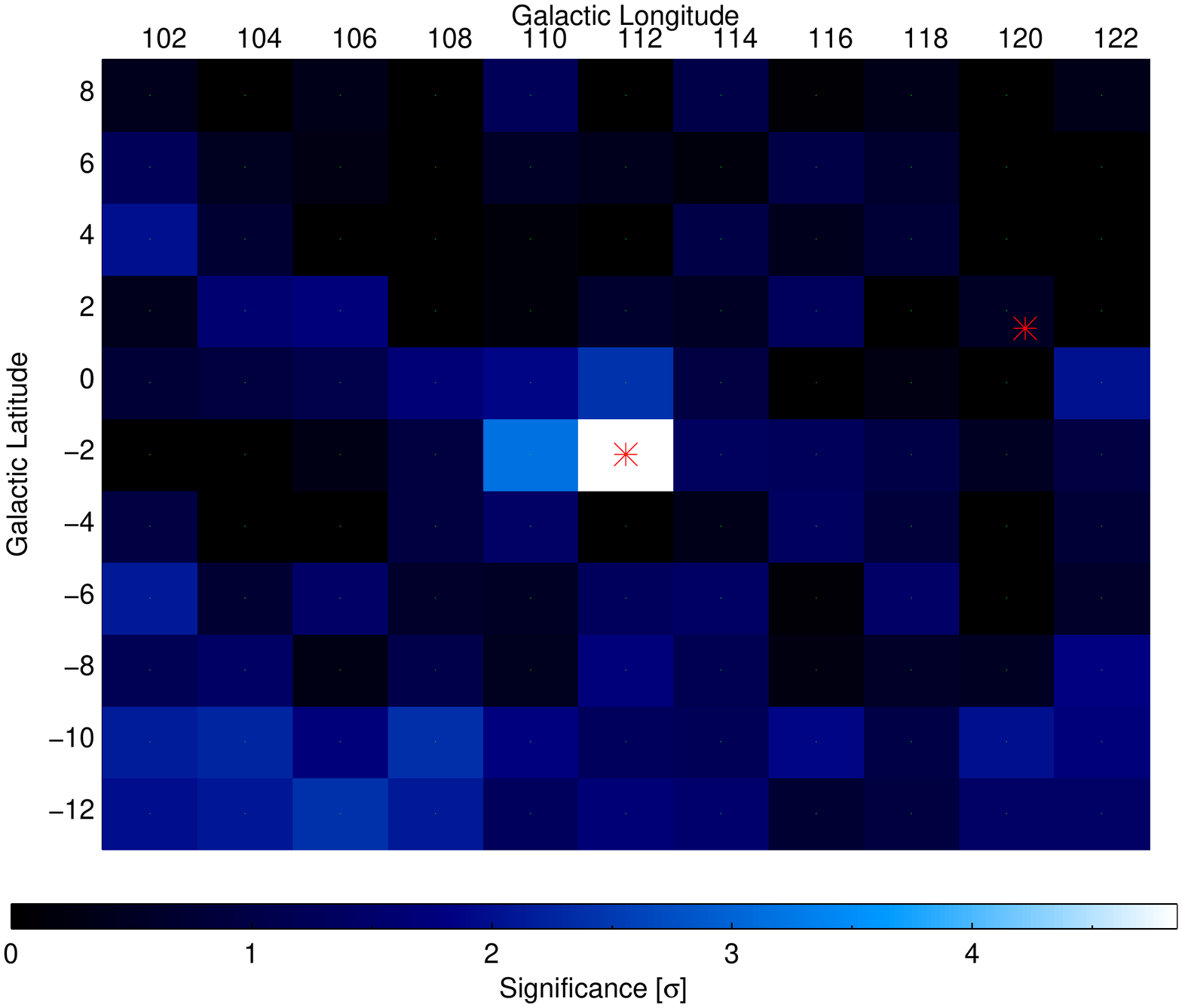}
        \includegraphics[width=\linewidth,trim={2.5cm 1.5cm 0.5cm 1.5cm}]{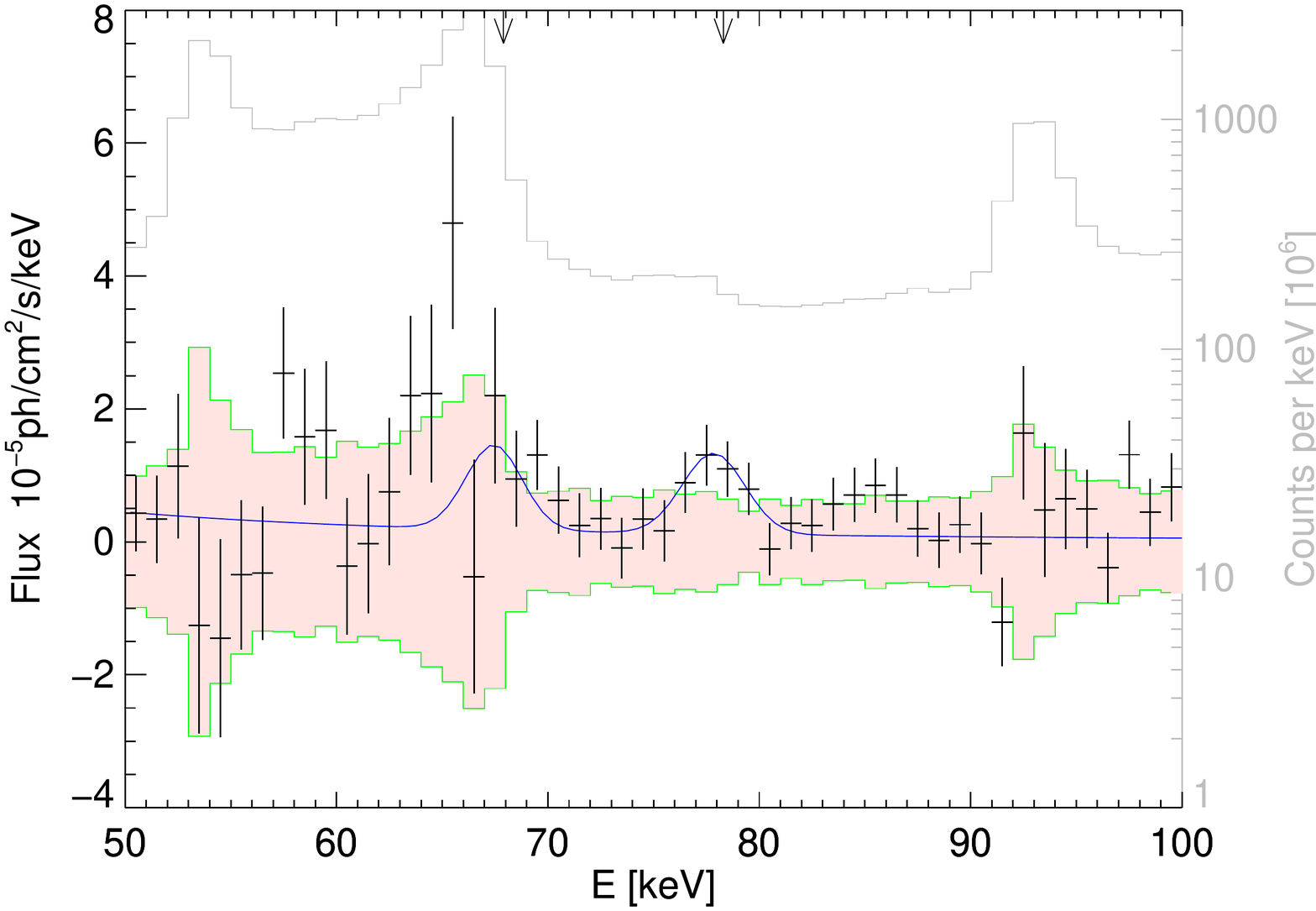}
        \caption{Upper: Significance map of $^{44}\rm Ti$ decay evaluated for the 78 and 1157\,keV line with color-coding in significance levels. Excess at the $4.9\sigma$ level is only found at the location of Cassiopeia A (center). The second red cross marks the location of Tycho's supernova remnant. Lower: The spectrum of Cassiopeia A is shown in black. The red shaded area contains the uncertainty band from the spectra obtained in a $20^{\circ}\times20^{\circ}$ area centered at Cassiopeia A. The gray spectrum shows the mission integrated background. It is evident that the uncertainty scales with the strength of the background leading to an increased uncertainty in the region 50-65\,keV. Arrows indicate the laboratory-determined centroid energies of both $^{44}\rm Ti$ decay lines. Clustered excess over the uncertainty band is found in the vicinity of the 78\,keV line. Unfortunately, the 68\,keV emission is located at the edge of a strong background line complex allowing for only a marginal detection probability of the line.}
  \label{fig:CasA_Detection}
\end{figure}

We used all available data for Cassiopeia A up to 2019\,AD, containing a total exposure of 11.2\,Ms for our analysis. Figure \ref{CasALELines} shows the spectrum of Cassiopeia A in the energy ranges of interest. The average reduced $\chi^2$ per fit energy bin is 1.001 ($\chi^2/dof$ = 92,772/92,658). We adopted a uniform power law underlying the line emission across the entire energy range between 30\,keV and 1200\,keV with a fit power-law index of $\alpha = -2.6 \pm 0.4$. The line signal with the highest significance for a single Gaussian shaped line is found for the 78\,keV line, with a significance of 3.6$\sigma$. The strong background lines of germanium lead to relatively large flux variations in the energy range between $\approx 50-65\,\mathrm{keV, to}$ such an extent that the 68\,keV decay line is only marginally detectable.
\\
In order to validate our findings and to avoid spurious detection, we searched for an emission that could mimic $^{44}\rm Ti$ decay from any celestial point source in a $20^{\circ}\times20^{\circ}$ area centered at Cassiopeia A. Possible emission was determined at locations in a spherical square-shaped grid, where each point is separated by 2$^{\circ}$ from the adjacent point, yielding a total of 121 grid points. Figure \ref{fig:CasA_Detection} shows the map of source significances for a 78+1157\,keV signal. Detection at $4.9\sigma$ is only found at the location consistent with Cassiopeia A, while other test points do not show significant signals. We additionally show (Fig. \ref{fig:CasA_Detection}) the $1\sigma$ uncertainty band determined form these 121 source locations, which, if interpreted as statistical signal fluctuations of a zero signal, bracket the spectrum of Cassiopeia A. A significant excess above the uncertainty band is clearly visible at the 78\,keV line, while the majority of the 68\,keV line lies within the statistical background-uncertainty band.
\begin{table*}\centering
\caption{Line shape parameters and derived quantities for the $^{44}\rm Sc$ and $^{44}\rm Ti$ decay in Cassiopeia A. Values are given for individual line fits and the combined 78 and 1157\,keV line fit, which is determined by fitting both lines with the same Doppler broadening and Doppler shift relative to their centroid energy. The last column contains the values for a combined fit of all three lines, but with Doppler parameters fixed to the values determined for the 78 + 1157\,keV line fit.}
\label{tab:CasAValues}
\begin{tabular}{l l l l l}
\hline\hline
                                                                                                                        & 78\,keV                                 & 1157\,keV                     & Combined                                                        & Three Lines Combined                                        \\ 
\hline 
Flux $[10^{-5}$\,ph\,cm$^{-2}$\,s$^{-1}$]                                       & 3.3$\pm$0.9                     & 9.5$\pm$3.0                   & 4.2$\pm$1.0                                                 & 4.6$\pm$0.8                                           \\
Mass  [$10^{-4}~\mathrm{M}_{\odot}$]                                    & 2.1$\pm$0.6                     & 5.9$\pm$1.9                   & 2.6$\pm$0.6                                                   & 2.9$\pm$0.5                                             \\ 
Centroid [keV]                                                                                                  & 77.7$\pm$0.5            & 1151$\pm$7.9          & 77.9$\pm$0.2; 1150.5$\pm$3.2          & 67.5 ; 77.9; 1150.5 (fixed)     \\
Shift [km\,s$^{-1}$]                                                                                    & $2400\pm1500$           & $1600\pm2000$         & $1800\pm800$                                          & 1800 (fixed)                                    \\
Line FWHM  [keV]                                                                                        & 2.3$\pm$0.8             & 40.0$\pm$6.7          & 2.4$\pm$0.9; 31.2$\pm$6.6                 & 2.3; 2.4; 31.2 (fixed)                                \\
Expansion Velocity [km\,s$^{-1}$]                                                       & 5500$\pm$2700           & 8900$\pm$1500         & 6400$\pm$1900                                         & 6400 (fixed)                                            \\
Significance [$\sigma$]                                                                         & 3.6                                     & 3.3                                   & 4.9                                                                     & 5.4                                                             \\              
\hline
\end{tabular}
\end{table*}
\\
The 78\,keV line, if represented by a Gaussian-shaped line, represents an integrated flux of $(3.3 \pm 0.9)\times10^{-5}$\,ph\,cm$^{-2}$\,s$^{-1}$ and a 3.6$\sigma$ detection level. With Eq. \eqref{eq:decaylaw}, this flux corresponds to an ejected $^{44}\rm Ti$ mass of $(2.1 \pm 0.6)\times10^{-4}$\,M$_{\odot,}$ for a remnant age of 330\,yr and a distance of 3300$\pm$100\,pc. The uncertainty is mainly due to the uncertainty in the determined flux, however, the uncertainty in the distance estimate is also incorporated in the result. The line is centered at $(77.7 \pm 0.5)\,\mathrm{keV}$, which is slightly red-shifted with respect to the laboratory-determined decay energy of 78.3\,keV \citep{Firestone2003}. This Doppler shift translates into a bulk motion of $(2400 \pm 1500)\,\mathrm{km\,s^{-1}}$ away from the observer. The full width at half maximum (FWHM) of the line is $(2.3 \pm 0.8)\,\mathrm{keV}$. This is broadened with respect to the instrumental resolution of 1.6\,keV FWHM at 78\,keV. We interpret this broadening of the line as Doppler broadening due to the expansion of the supernova remnant; this translates the FWHM of the line into an expansion velocity of $(5500 \pm 2700)\,\mathrm{km\,s^{-1}}$.
\\
Representing the 1157\,keV line with a simple Gaussian on top of the uniform power law, our best fit values suggest a significantly higher flux of $(9.5 \pm 3.0) \times 10^{-5}\,\mathrm{ph\,cm^{-2}\,s^{-1}}$. This flux can be overestimated due to the assumption of a single underlying continuum fit across the broad energy range. The continuum flux from our best fit power law (largely determined at energies below 100\,keV) in the energy region between 1125 -- 1175\,keV is $1.0 \pm 7.0 \times 10^{-7}\,\mathrm{ph\,cm^{-2}\,s^{-1}}$, consistent with zero. We estimated the potential offset in the fit continuum flux density by allowing a separate, constant offset in the energy range 1090 -- 1210\,keV, which accounts for a flux of $3.1 \pm 1.5 \times 10^{-5}\,\mathrm{ph\,cm^{-2}\,s^{-1}}$ in the energy region between 1125 -- 1175\,keV. Therefore, the line flux at 1157\,keV can be reduced to $(6.4 \pm 3.4) \times 10^{-5}\,\mathrm{ph\,cm^{-2}\,s^{-1}}$. The Gaussian is centered at $(1151 \pm 7.9)$\,keV, which is red shifted but compatible with the laboratory-determined energy of 1157\,keV. The line width is $(40.0 \pm 6.7)\,\mathrm{keV}$ FWHM. This corresponds to $(8900 \pm 1500)$\,km\,s$^{-1}$ expansion velocity.
\\
We find a combined signal with a significance of $4.9\sigma$ by simultaneously fitting two lines with identical Doppler shift, Doppler broadening, and integrated flux. The overall Doppler shift of the lines corresponds to a bulk ejecta velocity of $(1800 \pm 800)\,\mathrm{km\,s^{-1}}$. The Doppler broadening for both lines translates to $(6400 \pm 1900)\,\mathrm{km\,s^{-1}}$ expansion velocity, in agreement with values determined for the 78\,keV line alone. Due to the higher relative spectral resolving power of SPI at higher energies, the expansion velocity can be better constrained including the 1157\,keV line. The combined fit contains a flux of $(4.2 \pm 1.0) \times 10^{-5}\,\mathrm{ph\,cm^{-2}\,s^{-1}}$ per line. This higher flux corresponds to a $^{44}\rm Ti$ mass of $(2.6 \pm 0.6) \times 10^{-4}$\,M$_{\odot}$. Table \ref{tab:CasAValues} contains measured line parameters and derived physical quantities.
\\
Even though the uncertainties are high, we also include the 68\,keV line in our analysis. However, we adopted the kinematic values determined from the combined 78 and 1157\,keV line fit for this line, as the strong fluctuations induced by the strong background lines might lead to an artificial broadening of the line. The 68\,keV line is then observed with a single-line significance of 2.2$\sigma$. When linked to a common origin, the total significance for the three fit lines is then increased to $5.4\sigma$.

\subsection{SN 1987A}

SPI was pointed towards the LMC including SN 1987A for a total of 7\,Ms. With SPI's angular resolution of $\approx 2.7^\circ,$ we cannot distinguish between SN 1987A and other potential or known sources of high energy emission. In particular, the pulsar PSR B0540-69 and the high-mass X-ray binary LMC-X1 are located less than 1$^\circ$ apart from SN 1987A. We obtain an average reduced $\chi^2$ of 1.000 ($\chi^2/dof =$ 51,716/51,731) per energy bin. In Fig. \ref{SN1987A_spec}, we show the spectrum obtained from our data for a source located at the position of SN 1987A. We find no significant flux excess in either energy region that could be attributed to the decay chain emission of $^{44}\rm Ti$. \citet{Tueller1990} determined an expansion velocity of $3100\,\mathrm{km\,s^{-1}}$ from the line profiles of measured radioactive $^{56}\rm Co$. Assuming co-moving $^{44}\rm Ti$ ejecta, we adopted this value to determine $2\sigma$ upper limits on the flux. In our analysis, we searched for the combined signal of all three lines simultaneously, for which we determined a value of $1.8 \times 10^{-5}\,\mathrm{ph\,cm^{-2}\,s^{-1}}$ per line, corresponding to an upper mass limit of ejected $^{44}\rm Ti$ of $6.9 \times 10^{-4}\,\mathrm{M_{\odot}}$ for a distance of 49.6\,kpc, and a remnant age of 24\,yr. No systematically increased flux is observed in the 1157\,keV line with respect to the lines at 68 and 78\,keV.

\subsection{Vela Junior}
 
We modeled the supernova remnant as a source of extended emission with a 2D Gaussian emission profile and a diameter of 0.6$^{\circ}$ for the width of the remnant. This means that the 2$^{\circ}$ diameter contains $\approx 90\%$ of the expected $^{44}\rm Ti$ signal. The obtained fit is satisfactory with a reduced $\chi^2$ of 0.998 ($\chi^2/dof =$ 65,423/65,535) per energy bin. We find no signal for the decay of $^{44}\rm Ti$ (Fig. \ref{fig:VelaSN_remnant_spec}). We determined a 2$\sigma$ upper limit of $2.1 \times 10^{-5}\,\mathrm{ph\,cm^{-2}\,s^{-1}}$ for the combined signal of all three lines, assuming no bulk motion and an expansion velocity of $3000\,\mathrm{km\,s^{-1}}$. This corresponds to an upper limit for the ejected $^{44}\rm Ti$ mass of $3.3 \times 10^{-5}\,\mathrm{M_{\odot}}$ for the remnant age and distance of 690\,yr and 200\,pc, respectively. Considering the updated age and distance estimates of 2.4 -- 5.1\,kyr and $700\pm200$\,pc \citep{Allen2014}, the ejecta mass limits determined from our results significantly increase to a value $\leq 2.2 \times 10^{-1}\,\mathrm{M_{\odot}}$ for the lower age limit of 2.4\,kyr.

\subsection{Tycho's supernova remnant}
SPI was pointed towards the region containing Tycho's supernova remnant for a total of 10\,Ms. Figures \ref{fig:Ia_Low} and \ref{Figure_Ia_High} show the spectra obtained for Tycho in both energy regions relevant for our $^{44}\rm Ti$ search. The average reduced $\chi^2$ is 0.997 ($\chi^2/dof$ = 79,982/80,196) per energy bin. We find no significant excess for the emission in the three lines of the $^{44}\rm Ti$ decay chain. To determine our upper limits, we adopted an expansion velocity of $5000\,\mathrm{km\,s^{-1}}$. This value is in agreement with the expansion velocities found in the central ejecta (\cite{Sato2017}. We determined a 2$\sigma$ upper limit of $1.4 \times 10^{-5}\,\mathrm{ph\,cm^{-2}\,s^{-1}}$ for each line in the $^{44}\rm Ti$ decay, corresponding to an $^{44}\rm Ti$ ejecta-mass limit of  $4.8 \times 10^{-4}\,\mathrm{M_{\odot}}$ for a distance of 4.1\,kpc and a remnant age of 438\,yr.

\subsection{G1.9+0.3}

Due to its location close to the Galactic center, several hard X-ray sources \citep{Bird2016} are present within the 2.7$^{\circ}$ PSF of SPI around the position of G1.9+0.3. No signature for the decay of $^{44}\rm Ti$ is expected for these other sources, so  it appears a safe assumption that potential flux excess in the 68 and 78\,keV regime can be attributed to the emission from G1.9+0.3. The average reduced $\chi^2$ is 1.011 ($\chi^2/dof =$ 318,356/315,005). We interpret this as due to the possible presence of unresolved sources in SPI's field of view in the Galactic central region. The resulting spectra (Figs. \ref{fig:Ia_Low} and \ref{Figure_Ia_High}) show an underlying continuum from the spatially coincident sources, upon which we search for the imprints of the three decay lines.
\\
We find no significant excess in both energy ranges, determining a corresponding upper limit of $1.0 \times 10^{-5}\,\mathrm{ph\,cm^{-2}\,s^{-1}}$ for an assumed expansion velocity of $5000\,\mathrm{km\,s^{-1}}$ for $^{44}\rm Ti$ containing ejecta. This translates into a $^{44}\rm Ti$ yield of $0.3 \times 10^{-4}\,\mathrm{M_{\odot}}$ for a remnant age of 120\,yr and a distance of 8.5\,kpc. Even though this velocity is lower than the expansion of the remnant's blast wave, we believe our assumption is plausible, as the distribution of the ejecta containing radioactive $^{44}\rm Ti$ is uncertain anyway and may consist of clumps as seen for Cassiopeia A. 

We determined velocity-dependent limits, which depend on the expected line width for expansion velocities between 0 and $15000\,\mathrm{km\,s^{-1}}$. We obtain limits in the range from $0.7$ to $1.5 \times 10^{-5}\,\mathrm{ph\,cm^{-2}\,s^{-1}}$, assuming the same Doppler velocities for all three lines.
\\

\subsection{Kepler's supernova remnant}

In our analysis, we do not find emission from Kepler's supernova remnant in the two energy bands (Figs. \ref{fig:Ia_Low} and \ref{Figure_Ia_High}). The (2$\sigma$) upper limit is $1.1 \times 10^{-5}\,\mathrm{ph\,cm^{-2}\,s^{-1}}$. For a remnant age of 406\,yr and distance of 5.1\,kpc, the flux limit corresponds to a $^{44}\rm Ti$ ejecta mass limit of $4.0 \times 10^{-4}\,\mathrm{M_{\odot}}$.

\section{Discussion}\label{sec:discussion}

\begin{figure}
        \includegraphics[width=\linewidth,trim={1.5cm 2.5cm 2.0cm 2.5cm}]{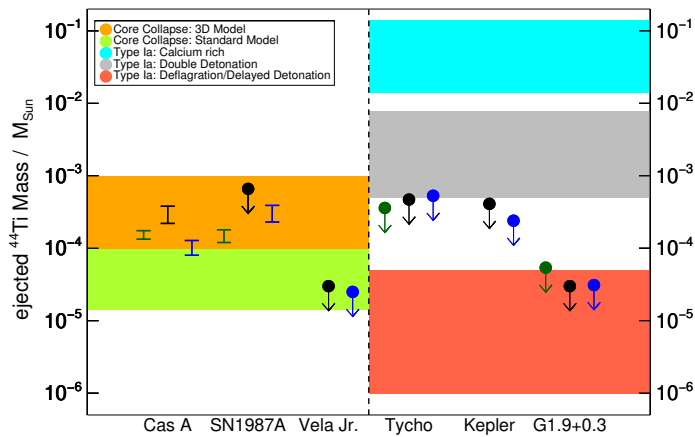}
        \caption{Left side compares upper limits and detection of Cassiopeia A with predictions made from core-collapse supernova models. Right hand side shows the predicted $^{44}\rm Ti$ ejecta yields for different type Ia supernova models. Upper limits obtained for the three youngest type Ia supernovae exclude the double detonation model and faint SN2005E-like scenarios. We include the detections of $^{44}\rm Ti$ in Cassiopeia A and SN1987 A with the NuSTAR and INTEGRAL/IBIS telescopes  in green and blue, respectively. Upper limits from further NuSTAR and IBIS observation are additionally included in green and blue (see Tab. \ref{table:summary} for references). Masses are determined for the distances in Tab. \ref{table:SN_remnant} and remnant age at the average observation date in the cited publications.}
  \label{Fig:ModelCompare}
\end{figure}

\subsection{Cassiopeia A}

For six analyzed supernova remnants, we find significant detection only for Cassiopeia A, with an integrated flux of $(4.2 \pm 1.0) \times 10^{-5}\,\mathrm{ph\,cm^{-2}\,s^{-1}}$ corresponding to an $^{44}\rm Ti$ ejecta mass of $(2.6 \pm 0.6) \times 10^{-4}\,\mathrm{M_{\odot}}$. 
\\
Conventional models of core-collapse supernova explosions \citep{Timmes1996,Magkotsios2010}, including models specific to the progenitor evolution of Cassiopeia A \citep{Young2006}, suggest $^{44}\rm Ti$ ejecta of less than $1.0 \times 10^{-4}\,\mathrm{M_{\odot}}$, significantly lower than the amount that we determine for Cassiopeia A. This underproduction of $^{44}\rm Ti$ in models is also supported by measurements from other instruments, consistently showing higher ejecta mass for Cassiopeia A \citep[$(2.4 \pm 0.9) \times 10^{-4}\,\mathrm{M_{\odot}}$][]{Siegert2015} and \citep[$(1.5 \pm 0.2) \times 10^{-4}\,\mathrm{M_{\odot}}$][]{Grefenstette2017} than the values from models of $\leq 1.0 \times 10^{-4}$\,M$_{\odot}$ obtained for a $30\,\mathrm{M_{\odot}}$ star \citep{Timmes1996,Limongi2018}. Figure \ref{Fig:ModelCompare} shows the expected yield of $^{44}\rm Ti$ for various supernova scenarios. The green shaded area represents standard, mostly piston-driven\footnote{Models that artificially inject the energy necessary for explosion} explosion models, in which nucleosynthesis is calculated by post-processing from the modeled thermodynamic evolution of the remnant. \citet{Harris2017} point out that including nucleosynthesis networks into the simulation, rather than post processing yields, can change the production of alpha nuclei, especially at intermediate mass range A = 36 -- 52, by an order of magnitude. The high $^{44}\rm Ti$ mass seen in Cassiopeia A shows that a more detailed treatment of explosive nucleosynthesis appears necessary. The 3D model of \citet{Wongwathanarat2017}, representing the special case of Cassiopeia A, suggests higher $^{44}\rm Ti$ masses, especially considering the clumpy and asymmetric distribution in the supernova remnant. The measured expansion velocity $(6400 \pm 1900)\,\mathrm{km\,s^{-1}}$ is compatible with models including Rayleigh-Taylor instabilities that lead to large-scale mixing of the inner ejecta with overlying stellar shells in type II-b supernova models \citep{Nomoto1995}. We determined a bulk motion of $(2200 \pm 1300)\,\mathrm{km\,s^{-1}}$. This suggests that the bulk of the ejecta is receding from the observer. Both the kinematics and the ejected mass of $^{44}\rm Ti$ support the interpretation that Cassiopeia A is an asymmetric supernova explosion. While we determined kinematic constraints from a spectral analysis, other evidence for an asymmetric explosion is provided from the spatially resolved analysis of the remnant with the NuSTAR telescope. \citet{Grefenstette2017} have found that the majority of the $^{44}\rm Ti$ -containing ejecta is expelled in a large solid angle, where the bulk of the ejecta moves away from the observer. Our measurements and the resulting velocity spread, determined from the Doppler broadening, is consistent with NuSTAR findings, which suggest a clumped nature of $^{44}\rm Ti$ -containing ejecta \citep{Grefenstette2017}. Despite concurring kinematic constraints, we determine a higher integrated flux in the $^{44}\rm Ti$ decay lines. Due to the different angular resolutions of the NuSTAR telescope \citep[18'' FWHM][]{Harrison2013} and the SPI spectrometer \citep[2.7$^{\circ}$ FWHM][]{Vedrenne2003}, different spherical surface areas for the integration of the flux are considered in both analyses. \citet{Grefenstette2017} considered the flux of an integrated emission from a region of 120'' radius centered on Cassiopeia A, containing all spatial points in which emission from $^{44}\rm Ti$ decay is detected in their analysis, and they gave upper limits on regions outside the 120'' radius. X-ray measurements both suggest a forward shock radius of 153" \citep{Gotthelf2001} and the presence of iron at radii between 110" -- 170" \citep{Willingale2002}. Co-moving $^{44}\rm Ti$ can be present at large radii extending as far outwards as the observed iron distribution. Flux from these unresolved regions contributes to the total flux of the $^{44}\rm Ti$ emission. Within the SPI's angular resolution, the entire surface area of the remnant is included, constituting the increased integrated flux we measured in our analysis. The increased line-of-sight beam width in SPI of 2.7$^{\circ}$ FWHM also includes areas outside of the supernova remnant. This also means that unresolved or previously unknown sources can contribute to the total flux in SPI measurements.
\\
For the first time, we also identified a very broad decay signature in the high-energy decay line at 1157\,keV, which also reveals the kinematic evolution of the supernova remnant. We determine an expansion velocity of $(8900 \pm 1500)\,\mathrm{km\,s^{-1}}$ and a line that is not significantly redshifted with $(1600 \pm 2000)\,\mathrm{km\,s^{-1}}$. This line contains an integrated flux of (9.5$ \pm $3.0)$\times 10^{-5}\,\mathrm{ph\,cm^{-2}\,s^{-1}}$ ((6.4$ \pm $3.4)$\times 10^{-5}\,\mathrm{ph\,cm^{-2}\,s^{-1}}$). A systematic offset of the high-energy line has been also observed with COMPTEL \citep{Iyudin1999} and with earlier SPI data \citep{Siegert2015}, however with less exposure on Cassiopeia A. In contrast to our updated values, \citet{Siegert2015} provided a more constraining expansion velocity and an overall different kinematic behavior of $^{44}\rm Sc$ decay in comparison to the 78\,keV line for a line centered at $(1158.0 \pm 3.6)\,\mathrm{keV}$. As shown by \citet{Grefenstette2017} $^{44}$Ti is ejected in clumps in Cassiopeia A. Each clump translates into a separate peak in the energy range between 1130 -- 1180\,keV, which blends into a broadened line. We described the entire emission by one Gaussian, which captures the overall expansion of the entire remnant. The line measured by \citet{Siegert2015} in the narrow energy window around 1157\,keV only captures parts, or one separate ejecta clump, of the total emission in the 1157\,keV regime. We find an increased flux in the 1157\,keV line, that cannot be explained by systematic effects alone. We speculate that the flux included in this line could be enhanced for different reasons: \textbf{1)} Excitation of the nuclear transition in $^{44}\rm Ca$, in addition to the decay of $^{44}\rm Sc$. Interaction with ambient material can lead to excitation of the nucleus. An excitation of the stable $^{44}\rm Ca$ nucleus by cosmic rays in the shock region of the supernova envelope might thus contribute to the flux in the 1157\,keV line. This mechanism would only influence the flux of the $^{44}\rm Ca^{*}$ line, as the half life of $^{44}\rm Sc$ is too short for efficient cosmic-ray-induced excitation. More analysis, in particular of other candidate nuclear de-excitation lines, is required to support this hypothesis. The most promising approach would be the detection of the de-excitation lines at 4.4\,MeV and 6.1\,MeV, which are the most prominent de-excitation lines caused by cosmic ray interaction in the shock front \citep{Summa2011}. A first search for these lines shows that flux values as high as those suggested by \citet{Summa2011} can be excluded. \textbf{2)} \citet{McKinnon2016} pointed out that two thirds of dust in the Milky Way-like galaxies can be produced by type II supernova events. The presence of dust grains composed of ejecta material in the vicinity of the supernova remnant or in the line of sight towards the supernova remnant can alter the observed flux ratios beyond the branching ratios. Attenuation coefficients for 68 and 78\,keV photons are higher than for the 1157\,keV line \citep{Iyudin2019} for common dust grain compositions. Including correction for branching ratios, we derived the following flux ratio:
\begin{equation}
        \frac{F_{78}}{F_{1157}} = 0.35 \pm 0.14\hspace{0.5cm}(0.52 \pm 0.32).
\end{equation}
This suggests that 16 -- 80$\%$ of the emission in the 78\,keV line could be absorbed by dust, located between INTEGRAL and Cassiopeia A.
\textbf{3)} The assumption of a Gaussian-shaped line for the total emission does not correctly represent the ejecta kinematics as found in Cassiopeia A. This can artificially lead to an increased flux in the 1157\,keV line. 

\subsection{SN 1987A}

We determine an upper limit of $1.8 \times 10^{-5}\,\mathrm{ph\,cm^{-2}\,s^{-1}}$ per decay line, assuming all lines share identical Doppler characteristics. We attribute this flux to the decay of $^{44}\rm Ti$, corresponding to an upper ejecta mass limit of $6.9 \times 10^{-4}\,\mathrm{M_{\odot}}$ for a remnant age of 24\,yr and a distance of 49.6\,kpc. We find no evidence for a systematically increased flux in the high energy line at 1157\,keV. Assuming that the $^{44}\rm Ti$ ejecta are contained in the central region of the supernova, the expansion velocity of the $^{44}\rm Ti$ ejecta should be lower than $1800 \,\mathrm{km\,s^{-1}}$ \citep{McCray2017}. In addition to an upper limit determined from the combined three lines, we give upper limits only for the 1157\,keV line for expansion velocities corresponding to the interior of the supernova core ($\mathrm{v_{exp}} \leq 1800\,\mathrm{km\,s^{-1}}$). For this velocity range, we determine flux limits that range between $(1.7 - 3.2) \times 10^{-5}\,\mathrm{ph\,cm^{-2}\,s^{-1}}$. The flux limit we derived for the $^{44}\rm Ti$ decay chain combined fit is compatible with direct detection of $^{44}\rm Ti$, as found by IBIS/INTEGRAL and NuSTAR \citep{Grebenev2012,Boggs2015}. Both analyses suggest narrow line broadening, which is compatible with slowly expanding $^{44}\rm Ti$ ejecta. Given the NuSTAR and IBIS/INTEGRAL fluxes, a significant offset in flux of the 1157\,keV line, either from less efficient absorption at higher energies or from an additional excitation, as seen in Cassiopeia A, should be detectable, albeit with small significance, within the SPI's sensitivity. 

\subsection{Vela Junior}

Vela Jr. still poses a mystery 20\,yr after the serendipitous detection of gamma-ray emission in the 1157\,keV $^{44}\rm Ca$ line by COMPTEL \citep{Iyudin1998}. Our upper limit of $2.1 \times 10^{-5}$\,ph\,cm$^{-2}$\,s$^{-1}$ assumes an extended source for the combined signal in all three decay lines, excluding a signal at the level reported by \citet{Iyudin1998} from COMPTEL data. Emission at the COMPTEL level was also excluded from the nondetection of the scandium fluorescence line \citet{Slane2001}. More recent studies with IBIS/INTEGRAL \citep{Tsygankov2016} found no excess in the energy bands of the 68 and 78\,keV lines, with an upper limit for the $^{44}\rm Ti$ flux of $1.8 \times 10^{-5}$\,ph\,cm$^{-2}$\,s$^{-1}$, which excludes the COMPTEL detection. \citet{Tsygankov2016} point out that they considered the remnant as a point-like source neglecting the apparent 2$^{\circ}$ diameter of the remnant. Measurements of the radial displacement in the northern rim of Vela Jr. suggest that the remnant age is $2.4 - 5.1\,\mathrm{kyr}$ \citep{Allen2014} at a distance of $0.5 - 1.0\,\mathrm{kpc}$, which is in contrast with the 0.7\,kyr age and 200\,pc distance estimates discussed by \citet{Iyudin1998}. Our results substantiate the higher age and larger distance.

\subsection{Tycho's supernova remnant}

Our upper limit of $1.4 \times 10^{-5}\,\mathrm{ph\,cm^{-2}\,s^{-1}}$ for the Tycho supernova remnant is in agreement with the upper limits determined with NuSTAR and INTEGRAL/IBIS \citep{Wang2014,Lopez2015} of $\geq 10^{-5}\,\mathrm{ph\,cm^{-2}\,s^{-1}}$ for moderate expansion velocities, with $^{44}\rm Ti$ spatially distributed over the entire remnant. However, detection has been claimed from observations with Swift/BAT \citep{Troja2014} at a flux of $1.3 \times 10^{-5}\,\mathrm{ph\,cm^{-2}\,s^{-1}}$ and $1.4 \times 10^{-5}\,\mathrm{ph\,cm^{-2}\,s^{-1}}$ for the 68 and 78\,keV lines, respectively. For an assumed distance of 4.1\,kpc, the upper limits correspond to an ejecta mass of the order of $10^{-4}\,\mathrm{M_{\odot}}$ of $^{44}\rm Ti$. \citet{Lopez2015} provided an ejecta-mass upper limit of $2.4 \times 10^{-4}\,\mathrm{M_{\odot}}$ of $^{44}\rm Ti$ for a distance of 2.3\,kpc. In all cases, we can exclude the double detonation and Ca-rich models as explosion scenario for this type Ia explosion remnant. The results are, however, in agreement with delayed detonation models. This model may be favored as it best reproduces the measured X-ray spectra from Tycho \citep{Badenes2006}. 

\subsection{G1.9+0.3}

We find no significant excess of $^{44}\mathrm{Ti}$ line emission at the position of the supernova remnant G1.9+0.3. We determine an upper flux limit of $1.0 \times 10^{-5}\,\mathrm{ph\,cm^{-2}\,s^{-1}}$. This translates into a mass limit of $0.3 \times 10^{-4}\,\mathrm{M_{\odot}}$ at a distance of 8.5\,kpc and a remnant age of 120\,yr, which excludes both double detonation and Ca-rich models for this candidate type Ia explosion. Predictions for classical delayed detonation models \citep{Maeda2010} suggest less than $10^{-5}$\,M$_{\odot}$ of $^{44}\rm Ti$, which would be in agreement with our results. We compare our results with the signal detected by \citet{Borkowski2010}. Using Chandra data, they reported a line at 4.1\,keV, which they attribute to a fluorescence transition in $^{44}\rm Sc$ following the electron capture decay on $^{44}\rm Ti$. Their inferred mass of $( 1 - 7) \times 10^{-5}\,\mathrm{M_{\odot}}$ of ejected $^{44}\rm Ti$ translates into a flux of $( 0.3 - 1.9) \times 10^{-5}\,\mathrm{ph\,cm^{-2}\,s^{-1}}$ in the decay lines at 68, 78, and 1157\,keV. Our upper limits ($0.7$ to $1.5 \times 10^{-5}\,\mathrm{ph\,cm^{-2}\,s^{-1}}$) are in conflict with the extrapolated fluxes ($( 0.3 - 1.9) \times 10^{-5}\,\mathrm{ph\,cm^{-2}\,s^{-1}}$) expected from the $^{44}\rm Sc$ fluorescence line for expansion velocity below $\sim$15000\,km\,s$^{-1}$. The discrepancy between the 4.1\,keV fluorescence line and the decay emission at hard X-ray energies suggests that the fluorescence line is not necessarily produced from the decay of $^{44}\rm Ti$ alone. Since the line is produced by the emission of a K$\alpha$ photon, which is independent of the scandium isotope, other stable isotopes (e.g., $^{45}\,\mathrm{Sc}$ ) co-produced in the supernova explosion can contribute to the fluorescence emission at 4.1\,keV.
\\
 \citet{Zoglauer2015} determined a 2$\sigma$ upper limit for the flux in the 68\,keV line of $1.5\times 10^{-5}$\,ph\,cm$^{-2}$\,s$^{-1}$ for a non-shifted line with a 4\,keV width (1$\sigma$), using measurements of the NuSTAR telescope. Results from the IBIS telescope \citep{Tsygankov2016} yield a 3$\sigma$ upper limit of  $9\times 10^{-6}$\,ph\,cm$^{-2}$\,s$^{-1}$. 

\subsection{Kepler's supernova remnant}

We determined a flux limit for Kepler of $1.1\times 10^{-5}$\,ph\,cm$^{-2}$\,s$^{-1}$. The derived mass limit disagrees with the double detonation and Ca-rich scenario; however, due to the uncertain distance to the remnant, ranging between 4.4\,kpc and 5.9\,kpc, the double detonation scenario cannot be explicitly excluded. Our result is in agreement with the nondetection of $^{44}\rm Ti$ emission with the COMPTEL and IBIS instrument \citep{Iyudin1999,Dupraz1997,Tsygankov2016}.

\subsection{Implications for supernova models}
The detection of $^{44}\rm Ti$ in only one Galactic supernova remnant provides a striking and significant conclusion regarding supernovae in the Galaxy. With an average core-collapse supernova rate of $\approx 1 - 3\,\mathrm{century}^{-1}$ \citep{Diehl2006,Bergh1991}, and current understanding of nucleosynthesis in supernovae, five supernova remnants with a $^{44}\rm Ti$ -decay line flux of more than $10^{-5}\,\mathrm{ph\,cm^{-2}\,s^{-1}}$ are expected to be visible in the Galaxy. Detection of a single remnant at the position of Cassiopeia A is unlikely, with a probability of less than 2.7\% \citep{The2006,Dufour2013}. With the high $^{44}\rm Ti$ mass measured in Cassiopeia A (and also SN 1987A), it is possible that the $^{44}$Ca content in the Galaxy is produced by a few, rare, $^{44}\rm Ti$ producing supernovae. It is possible that both Cassiopeia A and SN 1987A are the prototypes for asymmetric explosions producing a high ejecta $^{44}\rm Ti$ mass, whereas the majority of core-collapse supernovae explode in a more symmetric scenario, producing less $^{44}\rm Ti$ ejecta.\\

\begin{figure}
        \includegraphics[width=\linewidth,trim={1.5cm 1.5cm 2.0cm 2.5cm}]{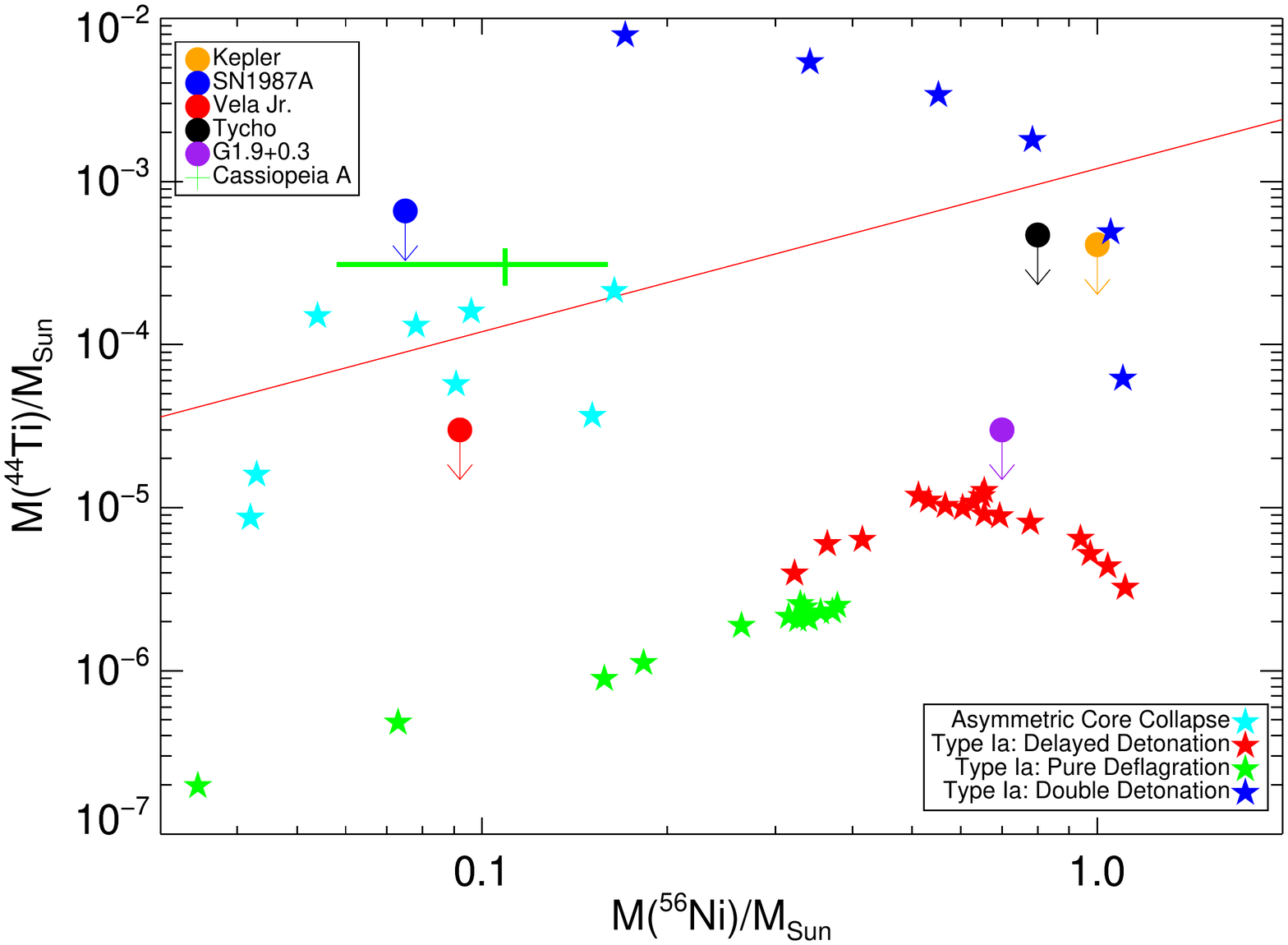}
        \caption{[$^{44}\rm Ti/^{56}\rm Ni$] ratio of our mass estimates and several supernova explosion models. The red line marks the solar [$^{44}\rm Ca /^{56}\rm Fe$] ratio \citep[$1.2\times 10^{-3}$;][]{Anders1989}, which is used as a reference criterion to judge supernova model subtypes. Ejected $^{56}\rm Ni$ masses for Cassiopeia A, SN 1987A, Tycho, Kepler and G1.9+0.3 are taken from \citep{Eriksen2009,Woosley1987,Badenes2006,Patnaude2012,Borkowski2013}, respectively. Vela Jr. $^{56}\rm Ni$ ejecta mass is modeled from a 25\,$\mathrm{M_{\odot}}$ star \citep{Maeda2003}. Model yields are from \citet{Wongwathanarat2017,Maeda2003,Seitenzahl2013,Fink2014,Fink2010}. }
  \label{Fig:CaFeRatio}
\end{figure}

We use the solar abundance value $[^{44}\mathrm{Ca}/^{56}\mathrm{Fe}]_\odot = 1.2\times 10^{-3}$ \citep[][]{Anders1989} of the $^{44}$Ca to $^{56}$Fe ratio, which are the end products of the $^{44}$Ti and $^{56}$Ni decay chain respectively, as another criterion to judge supernova model types, assuming that each of those supernova types would be the sole source of $^{44}$Ca and $^{56}$Fe as found in the Sun. Figure. \ref{Fig:CaFeRatio} shows the $^{44}$Ti to $^{56}$Ni ratio of the candidate sources in our analysis (dot symbols), together with modeled values for several supernova types (star symbols), and the solar $^{44}$Ca to $^{56}$Fe ratio (red line).\\

For Vela Jr., we adopted a $^{56}$Ni ejecta-mass estimate from the explosion model of a 25\,M$_\odot$ \citep{Maeda2003}. We note that the values for Vela Jr. may deviate from the solar ratio due to the assumed distance and age (200\,pc; 690\,yr), which underestimates the updated values determined by \citet{Allen2014}. The ejected $^{56}$Ni mass for Cassiopeia A is inferred from near-infrared spectral analysis \citep{Eriksen2009}. For SN 1987A, we used the $^{56}\rm Ni$ ejecta-mass estimate based on the early bolometric light curve by \citet{Woosley1987}. For type Ia supernova remnant candidates, $^{56}$Ni masses are obtained from \citet[][Tycho]{Badenes2006}, \citet[][Kepler]{Patnaude2012}, and \citet[][ G1.9+0.3]{Borkowski2013}, who inferred ejecta masses by comparing measured X-ray spectra with long-term simulated remnant models.\\

We infer from Fig. 5 that the measured $^{44}$Ti-to-$^{56}$Ni ratios of Cassiopeia A and SN 1987A are plausibly consistent with asymmetric core-collapse supernova models being responsible for the solar abundance ratio.\\
In contrast, the majority of type Ia supernova scenarios do not plausibly reproduce the measured solar $^{44}$Ca to $^{56}$Fe ratio. Type Ia supernovae appear to consistently produce a $^{56}\rm Ni$ ejecta mass between $10^{-1}-10^{0}$ M$_{\odot}$ \citep{Dhawan2016,Wang2008,Stritzinger2006}. In model calculations, nucleosynthesis yields of intermediate mass elements (such as $^{44}\rm Ti$) are highly dependent on physical conditions during nuclear burning. In centrally ignited, pure-deflagration, and delayed-detonation scenarios, nucleosynthesis occurs mainly in a high-density regime, producing only a small amount of $^{44}\rm Ti$, while the majority of nuclear fuel is completely burned to iron group elements \citep{Seitenzahl2013,Fink2014}. On the other hand, nucleosynthesis in double-detonation supernova scenarios also occurs during the initial burning of the surface helium layer at lower densities, allowing for the production of large amounts of $^{44}\rm Ti$ on the surface \citep{Fink2010,Sim2012}. Pure-deflagration and delayed-detonation type Ia supernovae (green and red star symbols, Fig.\ref{Fig:CaFeRatio}) therefore could not be major contributors to solar $^{44}$Ca, while double detonation supernovae (blue star symbols in Fig.\ref{Fig:CaFeRatio}) could. Our constraints for all three candidate remnants of type Ia supernovae show that they cannot be sole sources of the solar abundance of $^{44}$Ca and $^{56}$Fe together.\\

Under the assumption that core-collapse supernovae are, in general, responsible for the solar [$^{44}\rm Ca /^{56}\rm Fe]_{\odot}$ ratio, we can use our $^{44}\rm Ti$ limits and the average simulated $^{44}\rm Ti$ and $^{56}\rm Ni$ yields to constrain type Ia sub-type rates: We obtain a ratio $\geq 2.6:1$ of delayed detonation/deflagration to double detonation events, necessary to maintain and not violate the solar $[^{44}\rm Ca/^{56}\rm Fe]_\odot$ ratio.

\begin{table*}\centering
\caption{Values for the masses and fluxes of the six young supernova remnants. Fluxes are determined for the $^{44}\rm Sc$ and $^{44}\rm Ti$ decay separately and also with a combined fit for the the most stringent constraints in our analysis. We also include results determined with the NuSTAR and Integral/IBIS telescope, separated by the dashed line. References: (1) \citep{Grefenstette2017}; (2) \citep{Tsygankov2016}; (3) \citep{Boggs2015}; (4) \citep[two lines combined][]{Grebenev2012}; (5) \citep{Lopez2015}; (6) \citep{Wang2014}; and (7) \citep{Zoglauer2015}.}
\label{table:summary}
\begin{tabular}{c c c c c c c}
\hline\hline
                                                                                                                                & Cassiopeia A                    & SN 1987A                       & Vela Jr                 & Tycho                 & Kepler                        & G1.9+0.3        \\ 
\hline
Flux $^{44}\rm Sc$ $[10^{-5}$\,ph\,cm$^{-2}$\,s$^{-1}$]         & 9.5$\pm$3.0                   & <4.1                                    & <4.7                  & <6.2                   & <2.6                          & <3.7          \\ 
Flux $^{44}\rm Ti$ $[10^{-5}$\,ph\,cm$^{-2}$\,s$^{-1}$]         & 3.3$\pm$0.9                   & <1.9                            & <2.8                  & <1.5                  & <1.3                            & <1.1          \\ 
Flux Comb  $[10^{-5}$\,ph\,cm$^{-2}$\,s$^{-1}$]                         & 4.2$\pm$1.0             & <1.8                                  & <2.1                   & <1.4                  & <1.1                          & <1.0            \\ 
Mass Comb  [$10^{-4}$\,M$_{\odot}$]                                                     & 2.6$\pm$0.6             & <6.9                                  & <0.3                   & <4.8                  & <4.0                          & <0.3            \\ 
\hdashline
NuSTAR $[10^{-5}$\,ph\,cm$^{-2}$\,s$^{-1}$]                                     & 1.8$\pm$0.3 (1)         & 0.35$\pm$0.07 (3)     & \textendash   & <1.0 (5)              & \textendash             & <1.5 (7)      \\
Integral/IBIS $[10^{-5}$\,ph\,cm$^{-2}$\,s$^{-1}$]                      & 1.3$\pm$0.3 (2)         & 1.7$\pm$0.4 (4)               & <1.8 (2)              & <1.5 (6)                & <0.63 (2)             & <0.9 (2)      \\
\hline
\end{tabular}
\end{table*}
\subsection{Summary}
In this work, we searched for the signature of gamma rays produced in the decay chain of $^{44}\rm Ti$ in the six young nearby supernova remnants Cassiopeia A, SN 1987A, Vela Jr., Tycho's supernova, Kepler's supernova, and G1.9+0.3. In Tab. \ref{table:summary}, we list the mass estimates we derived from SPI/INTEGRAL data acquired over the entire mission duration of 17\,yr.
\\
We only detect emission in the supernova remnant Cassiopeia A. Inferred masses of more than $2-3 \times 10^{-4}$\,M$_{\odot}$ ejected $^{44}\rm Ti$ exceed theoretical predictions. Upper limits determined for Vela Jr. exclude the detection of the $^{44}\rm Sc$ decay line found with COMPTEL by \citet{Iyudin1999}. The detection of $^{44}\rm Ti$ hard X-ray lines in SN 1987A \citep{Grebenev2012,Boggs2015} cannot be confirmed. 
We exclude models predicting high yields of $^{44}\rm Ti$ such as the double-detonation models and Ca-rich explosion \citep{Waldman2011,Perets2010} model for the three Galactic thermonuclear supernova remnants G1.9+0.3, Tycho, and Kepler.

\begin{acknowledgements}
The INTEGRAL/SPI project has been completed under the responsibility and leadership of CNESS; we are grateful to ASI,CEA, CNES, DLR (Nos. 50OG 1101 and 1601), ESA, INTA, NASA and OSTC for support of this ESA space science mission. Thomas Siegert is supported by the German Research Society (DFG-Forschungsstipendium SI 2502/1-1)
\end{acknowledgements}

\bibliography{Bibliography}

\appendix
\section{Additional supernovae spectra}
\begin{figure}
        \includegraphics[width=\linewidth,trim={1.5cm 1.5cm 1.5cm 1.5cm}]{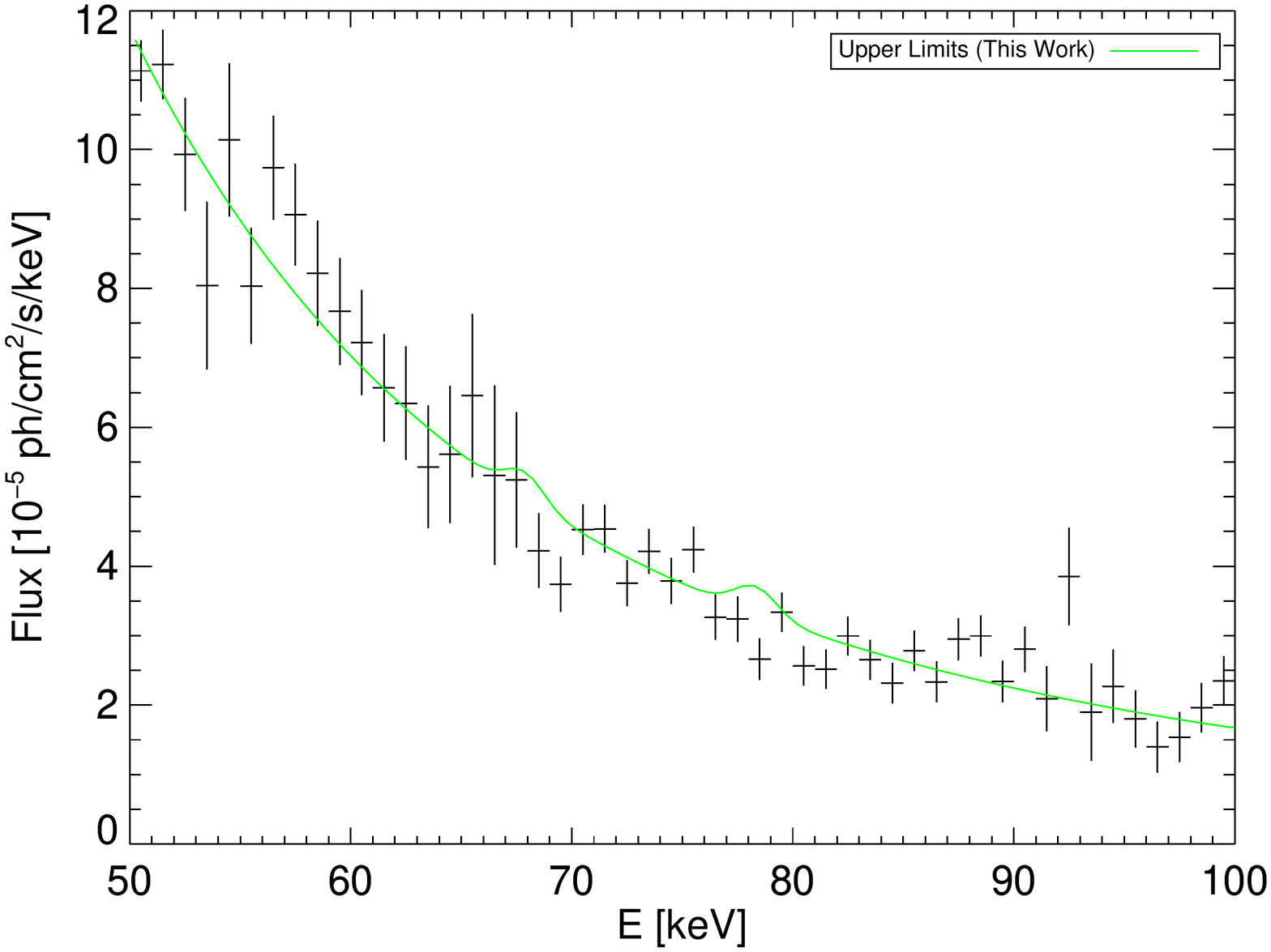}
        \includegraphics[width=\linewidth,trim={1.5cm 1.5cm 1.5cm 1.5cm}]{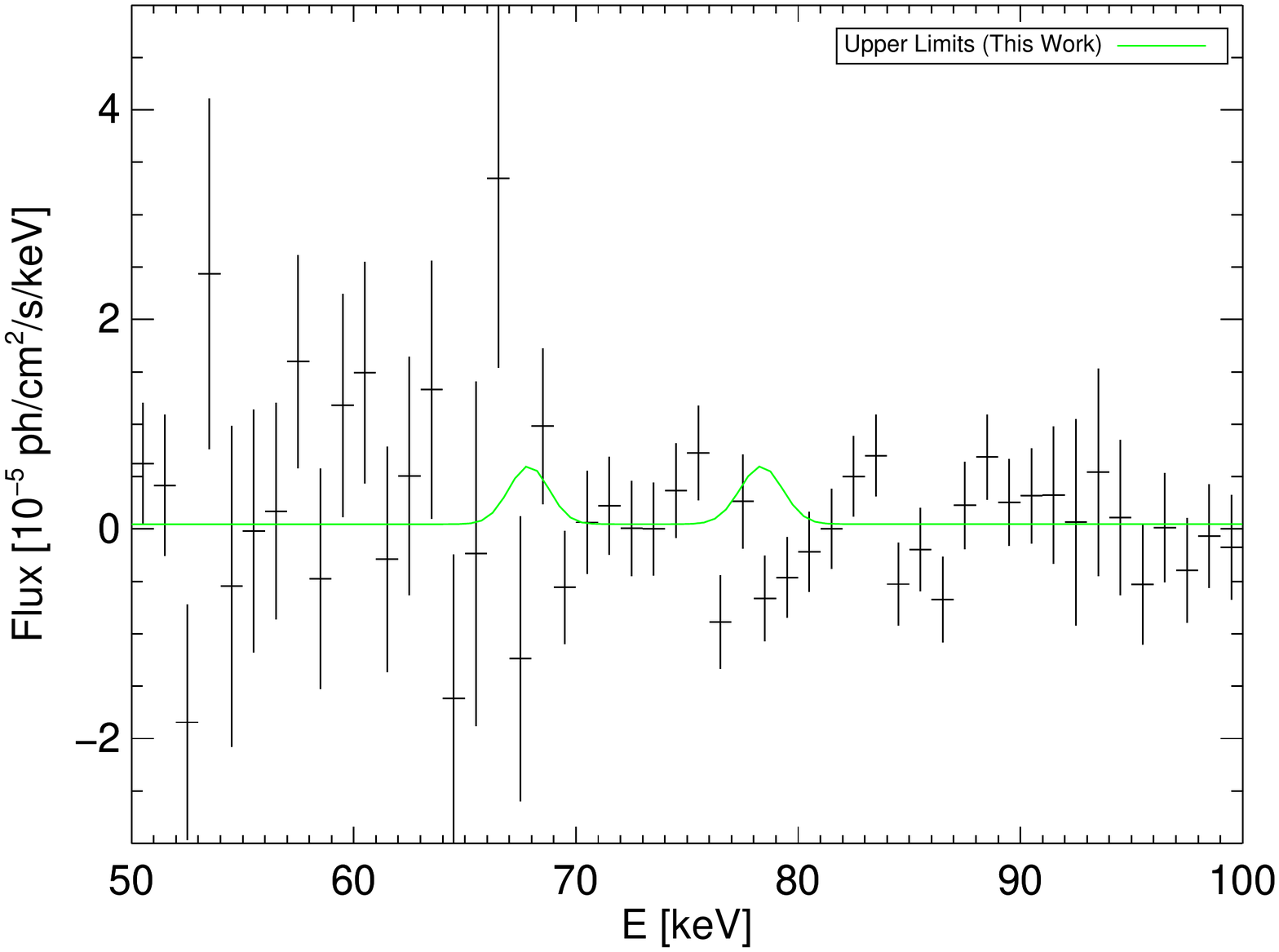}
        \includegraphics[width=\linewidth,trim={1.5cm 1.5cm 1.5cm 1.5cm}]{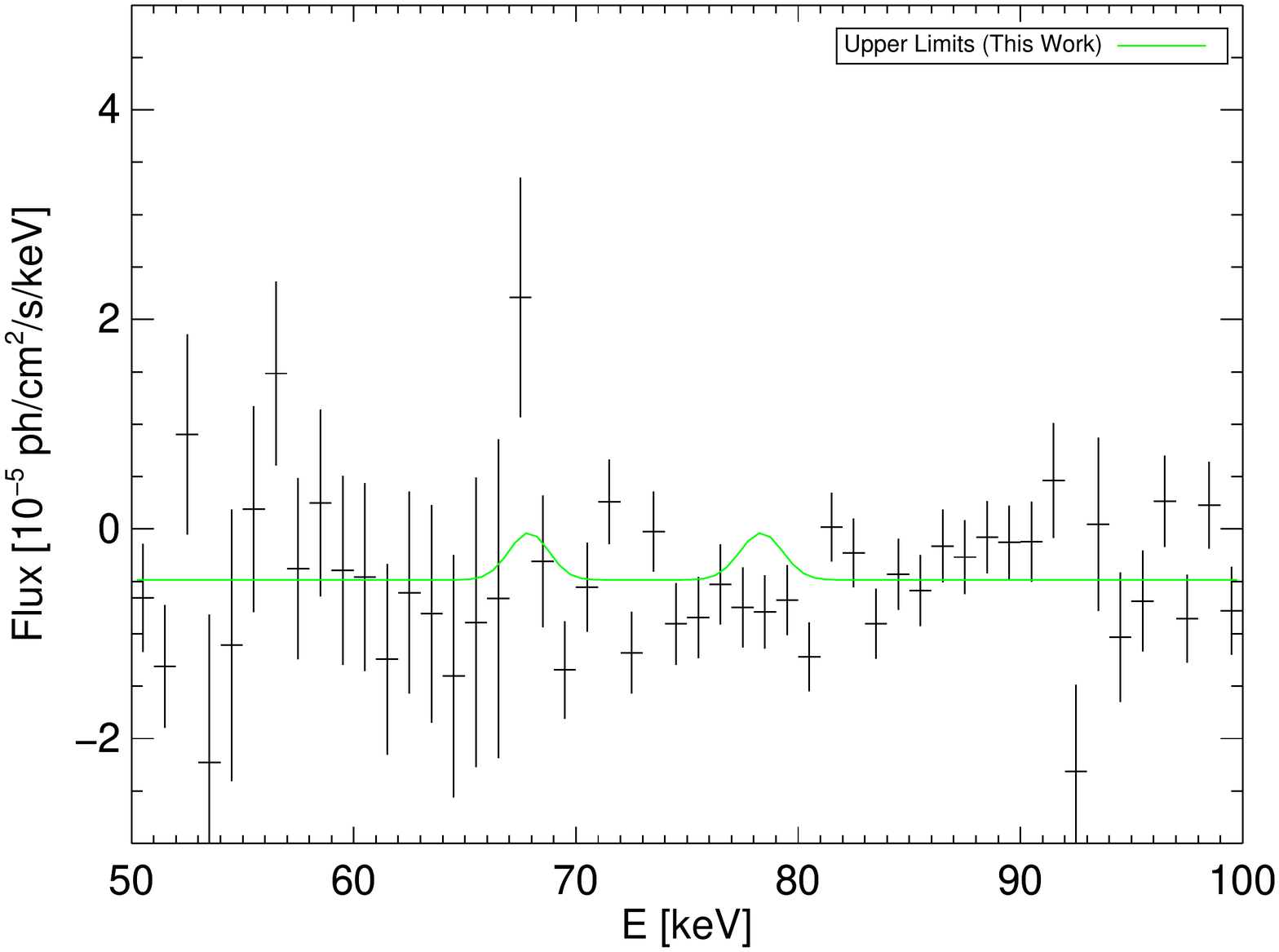}       
        \caption{Spectra containing the $2\sigma$ upper limits determined for the Type Ia supernovae. From top to bottom: G1.9+0.3, Tycho and Kepler.  }
  \label{fig:Ia_Low}
\end{figure}

\begin{figure}
        \includegraphics[width=\linewidth,trim={1.5cm 1.5cm 1.5cm 1.5cm}]{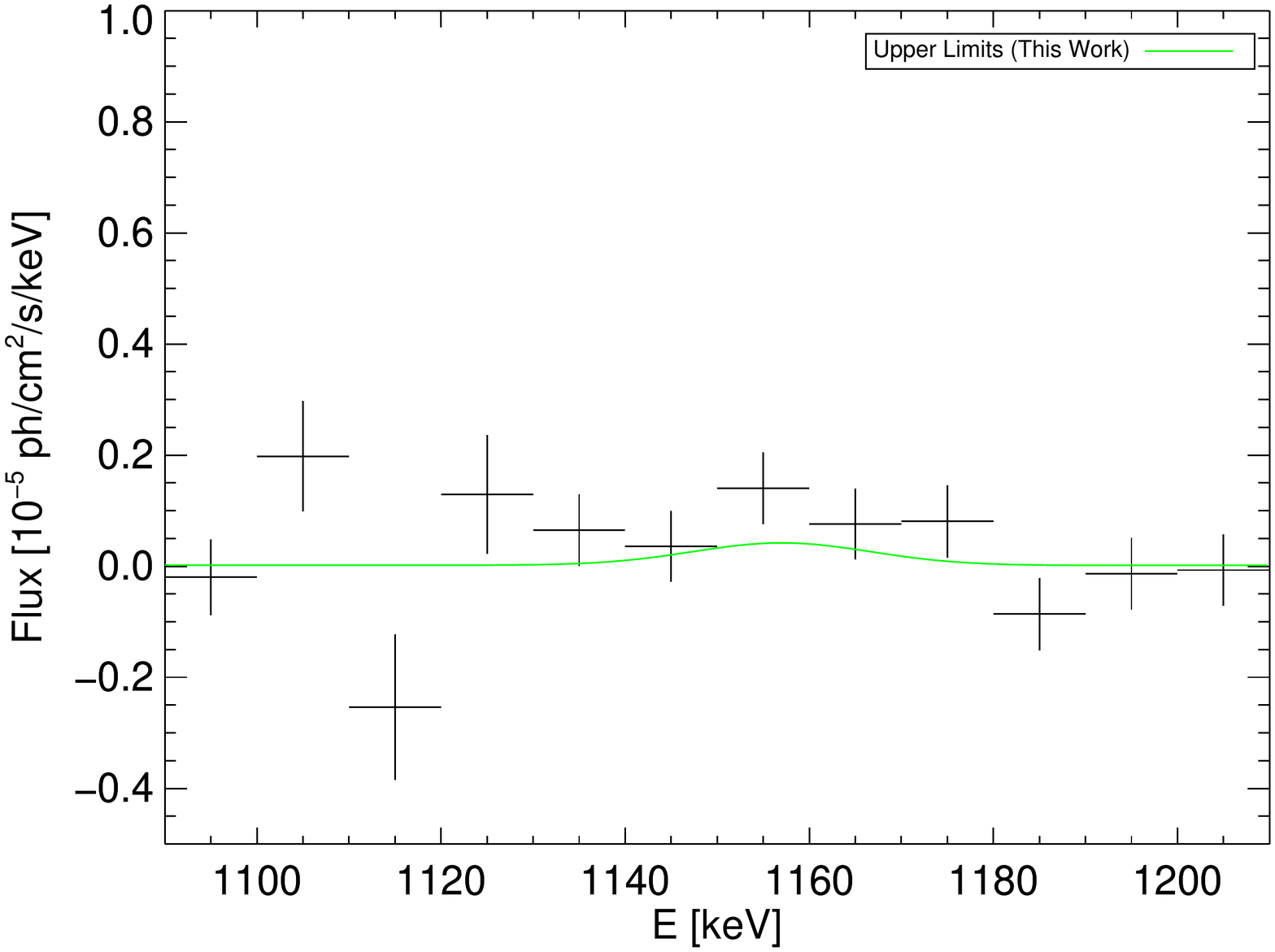}
        \includegraphics[width=\linewidth,trim={1.5cm 1.5cm 1.5cm 1.5cm}]{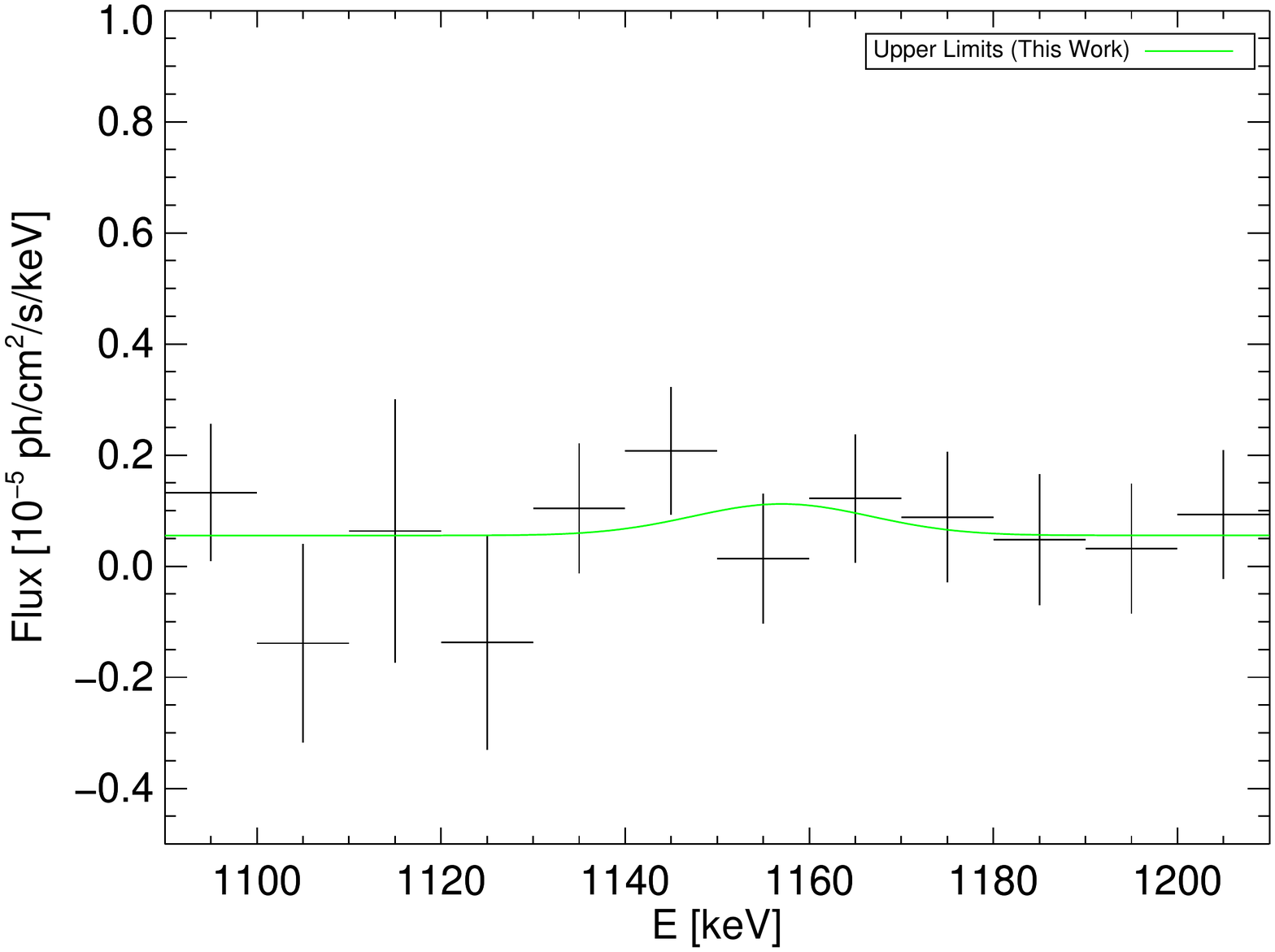}
        \includegraphics[width=\linewidth,trim={1.5cm 1.5cm 1.5cm 1.5cm}]{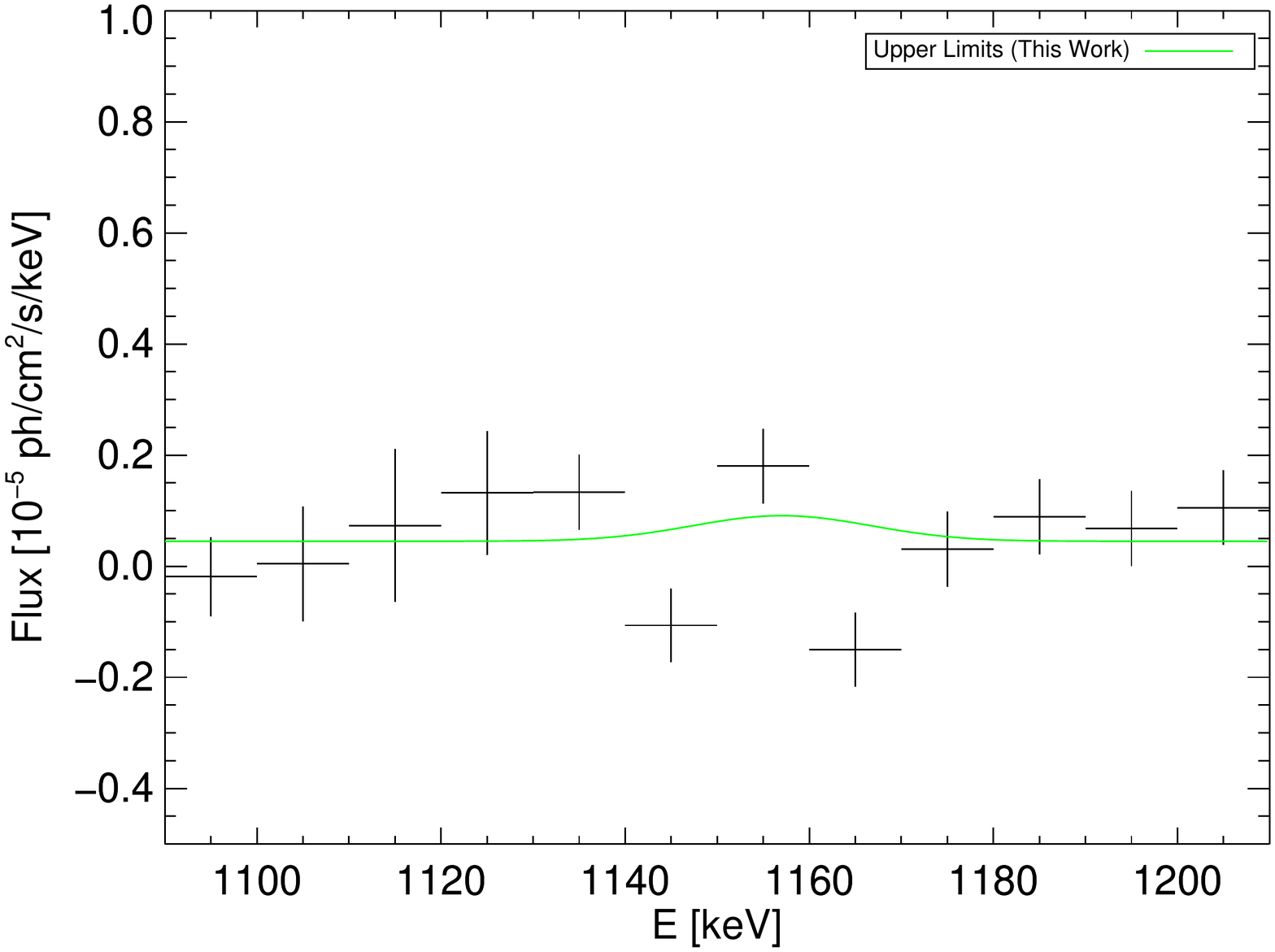}      
        \caption{Same as Fig. \ref{fig:Ia_Low} for the energy range containing the $^{44}\rm Sc$ decay line.}
  \label{Figure_Ia_High}
\end{figure}

\begin{figure}
        \includegraphics[width=\linewidth,trim={1.5cm 1.5cm 1.5cm 1.5cm}]{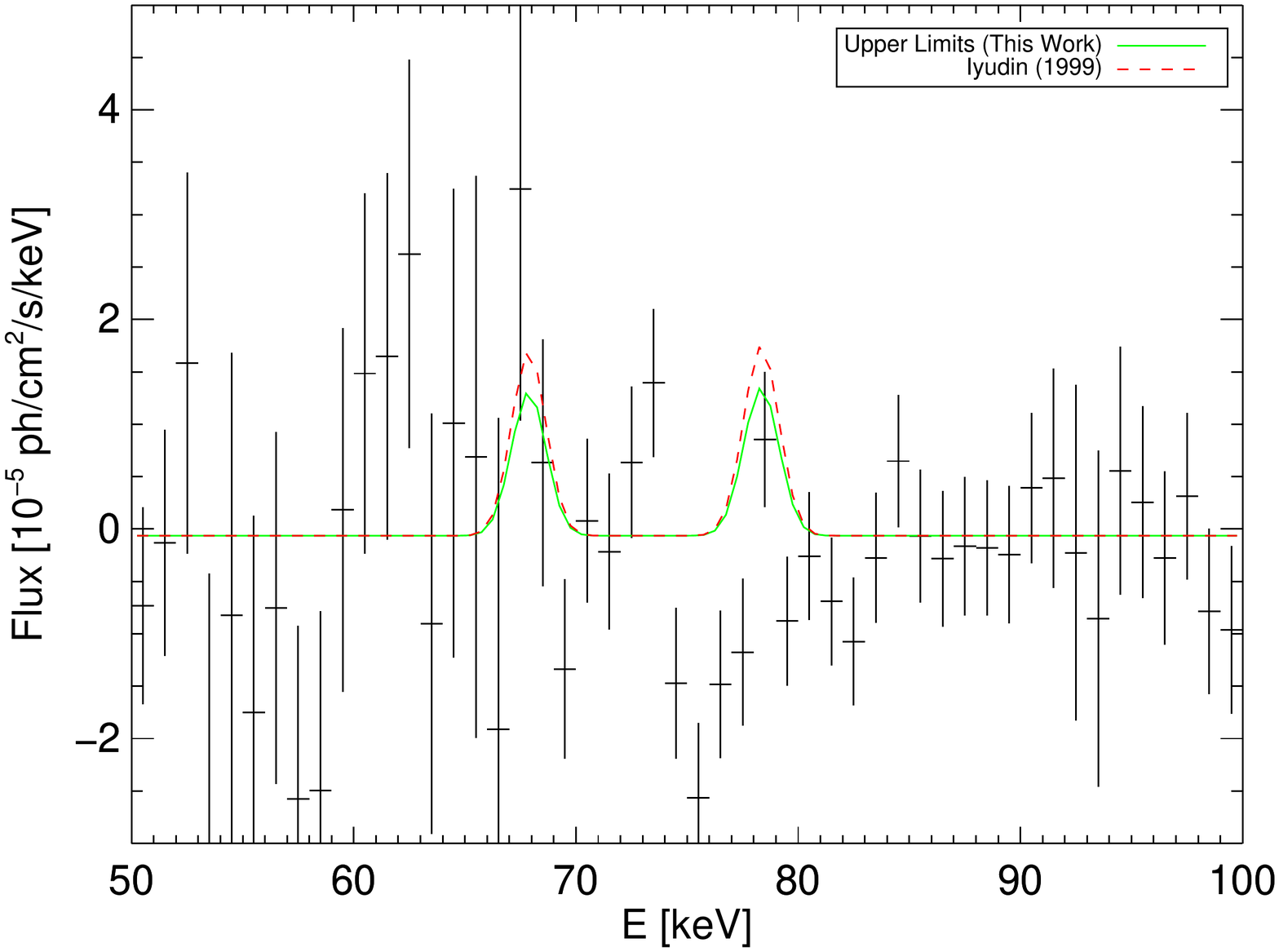}
        \includegraphics[width=\linewidth,trim={1.5cm 1.5cm 1.5cm 1.5cm}]{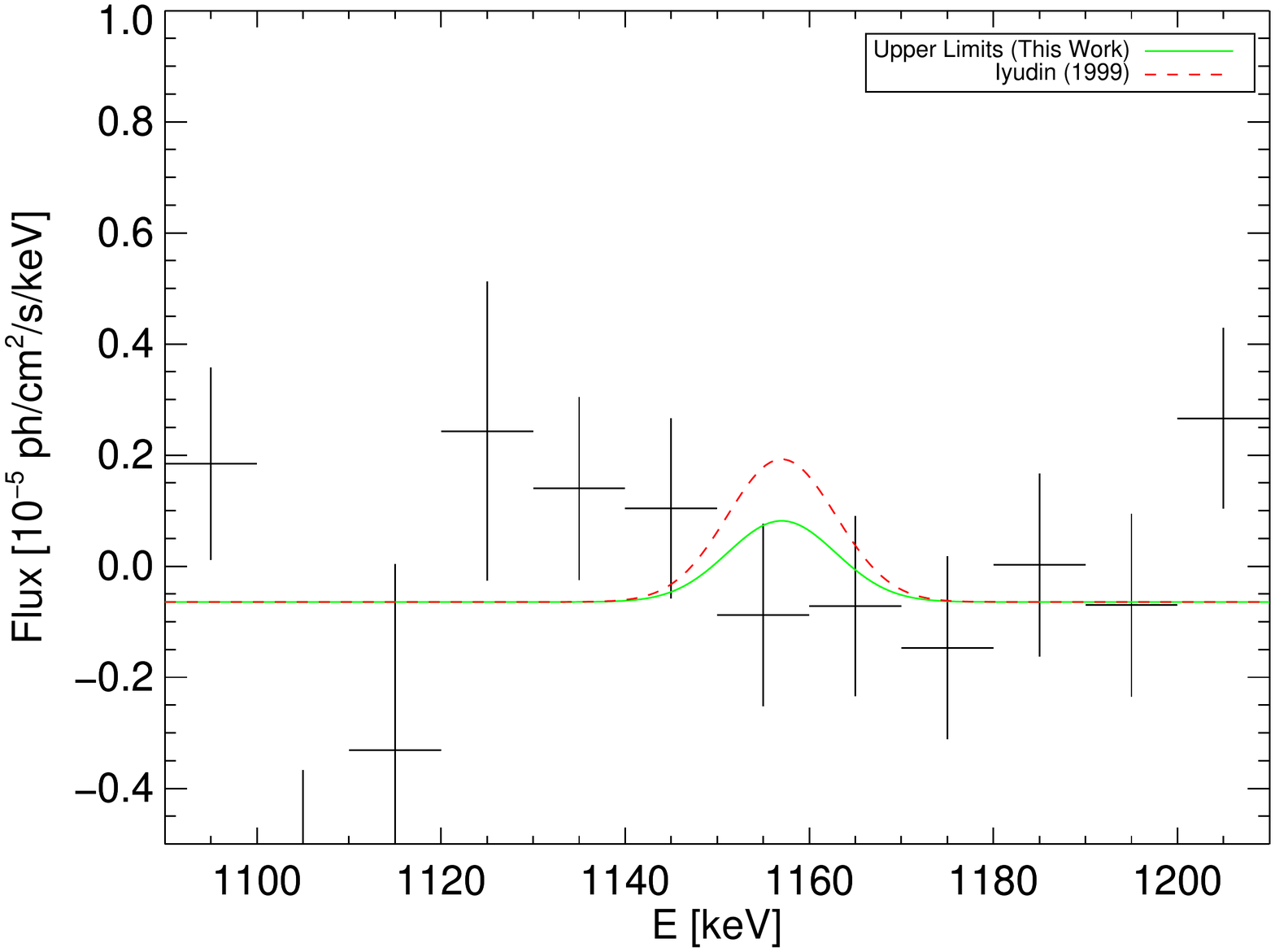}
        \caption{Spectrum of Vela Jr. in both energy regions. No significant excess in can be found in either of the two regions for all three lines. The remnant was modeled with a Gaussian-shaped emission region with 0.6$^{\circ}$ Gaussian width. Upper limits are shown in green. Red dashed line is the flux determined with COMPTEL data \citep{Iyudin1999}}
  \label{fig:VelaSN_remnant_spec}
\end{figure}

\begin{figure}
        \includegraphics[width=\linewidth,trim={1.5cm 1.5cm 1.5cm 1.5cm}]{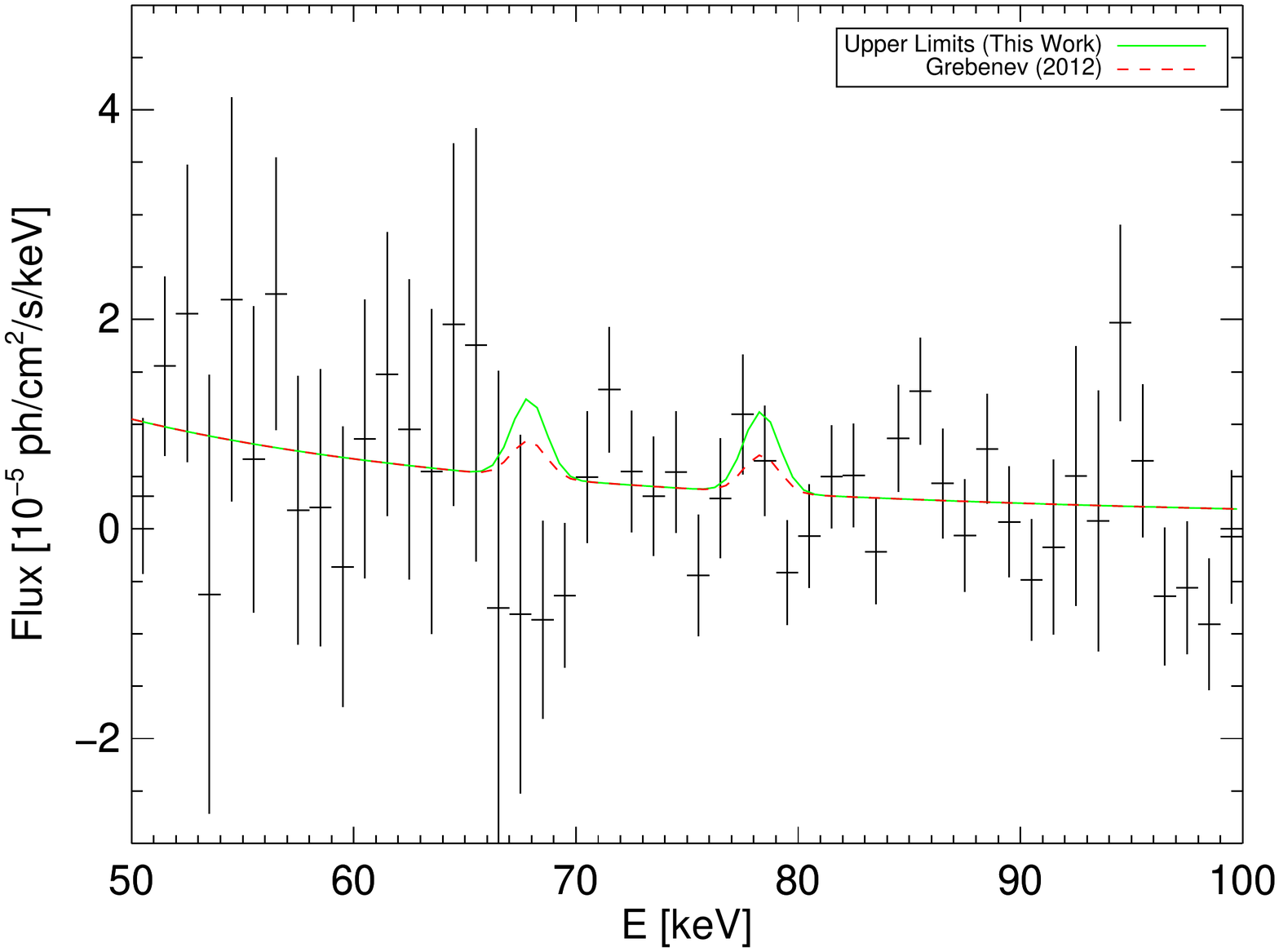}
        \includegraphics[width=\linewidth,trim={1.5cm 1.5cm 1.5cm 1.5cm}]{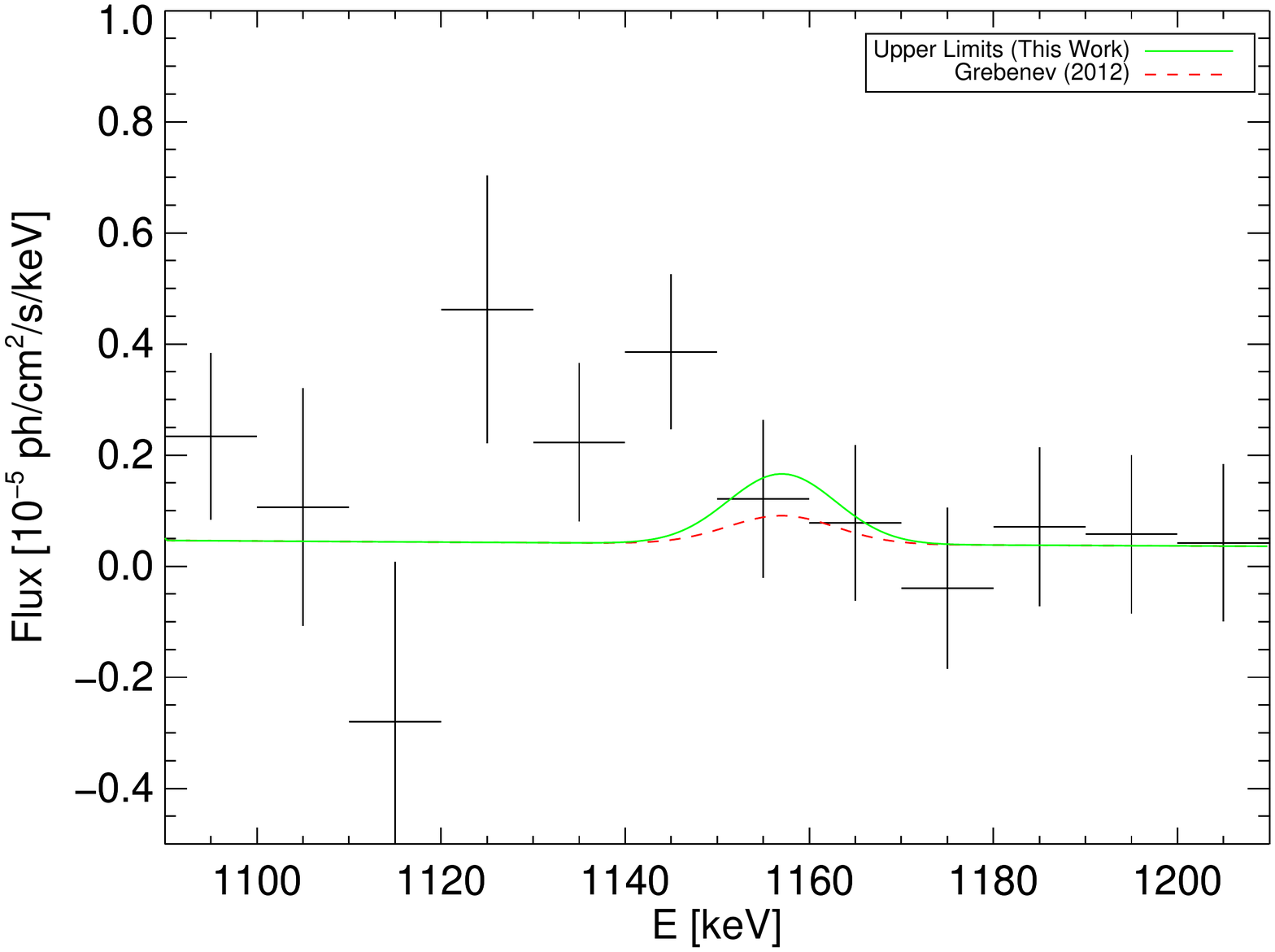}
        \caption{Upper panel: Spectrum in the energy range from 50-100\,keV for a source at celestial position of SN 1987A. The spectrum is fit with a power law accounting for the emission of LMC-X1 and PSR B0540-69, which are spatially indistinguishable from SN 1987A. The green line shows the power-law continuum and the 2$\sigma$ line limits. Lower panel: Same as upper panel but for energy range 1090-1210\,keV. We allow for a constant offset to account for possible diffuse emission sources located within the $2.7^\circ$ angular resolution of SPI. We do not find significant flux excess for any of the $^{44}\rm Ti$ decay lines. The $2\sigma$ upper-limit flux determined for an expansion velocity of $3000 $\,km\,s$^{-1}$ is $1.8 \times 10^{-5}\,\mathrm{ph\,cm^{-2}\,s^{-1}}$ corresponding to $7.0 \times 10^{-4}\,\mathrm{M_{\odot}}$ synthesized $^{44}\rm Ti$ (green line). Red dashed line corresponds to the IBIS flux \citep{Grebenev2012} as it would be seen in SPI data.}
  \label{SN1987A_spec}
\end{figure}

Figures \ref{fig:Ia_Low}, \ref{Figure_Ia_High}, \ref{fig:VelaSN_remnant_spec}, and \ref{SN1987A_spec} contain the spectra of the supernova remnants G1.9+0.3, Tycho, Kepler, Vela Jr., and SN 1987A not shown in the main text.

\section{Investigating systematics of analysis approach} \label{sec:BG_Append}
We tested our analysis method for consistency in two scenarios. We used data taken in the high latitude region close to the Galactic north pole and analyzed the two point sources NGC4388, for which we expected no emission in the $^{44}\rm Ti$ decay chain and an "empty" region with no known celestial source.
Both "sources" are close to the Galactic north pole with latitude 60$^\circ$ <b < 80$^\circ$ and longitude $-320^\circ$ < l < -270$^\circ$ containing a total of 11.5\,Ms exposure for the entire INTEGRAL mission time. This region contains only a few sources, making it ideal for testing the analysis approach in an "empty" celestial region. The sources can be outside the fully coded field of view for various observation times, reducing the exposure on specific locations. The image response function was not calculated for angles between the pointing direction and source location of more than 25 degree offset, and we assumed no source contribution in these cases.\begin{figure}
        \includegraphics[width=\linewidth,trim={1.5cm 1.5cm 1.5cm 1.5cm}]{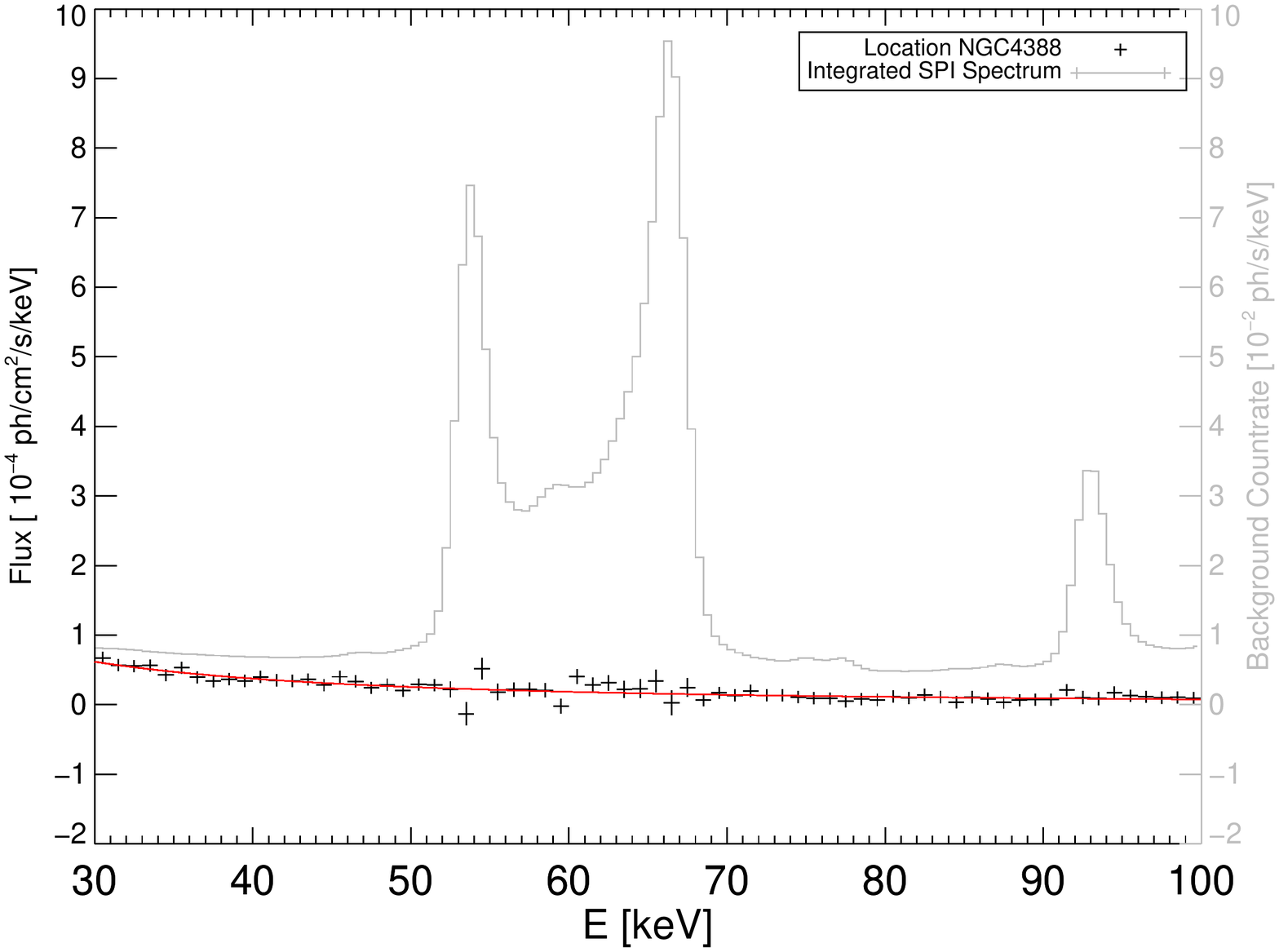}
        \includegraphics[width=\linewidth,trim={1.5cm 1.5cm 1.5cm 1.5cm}]{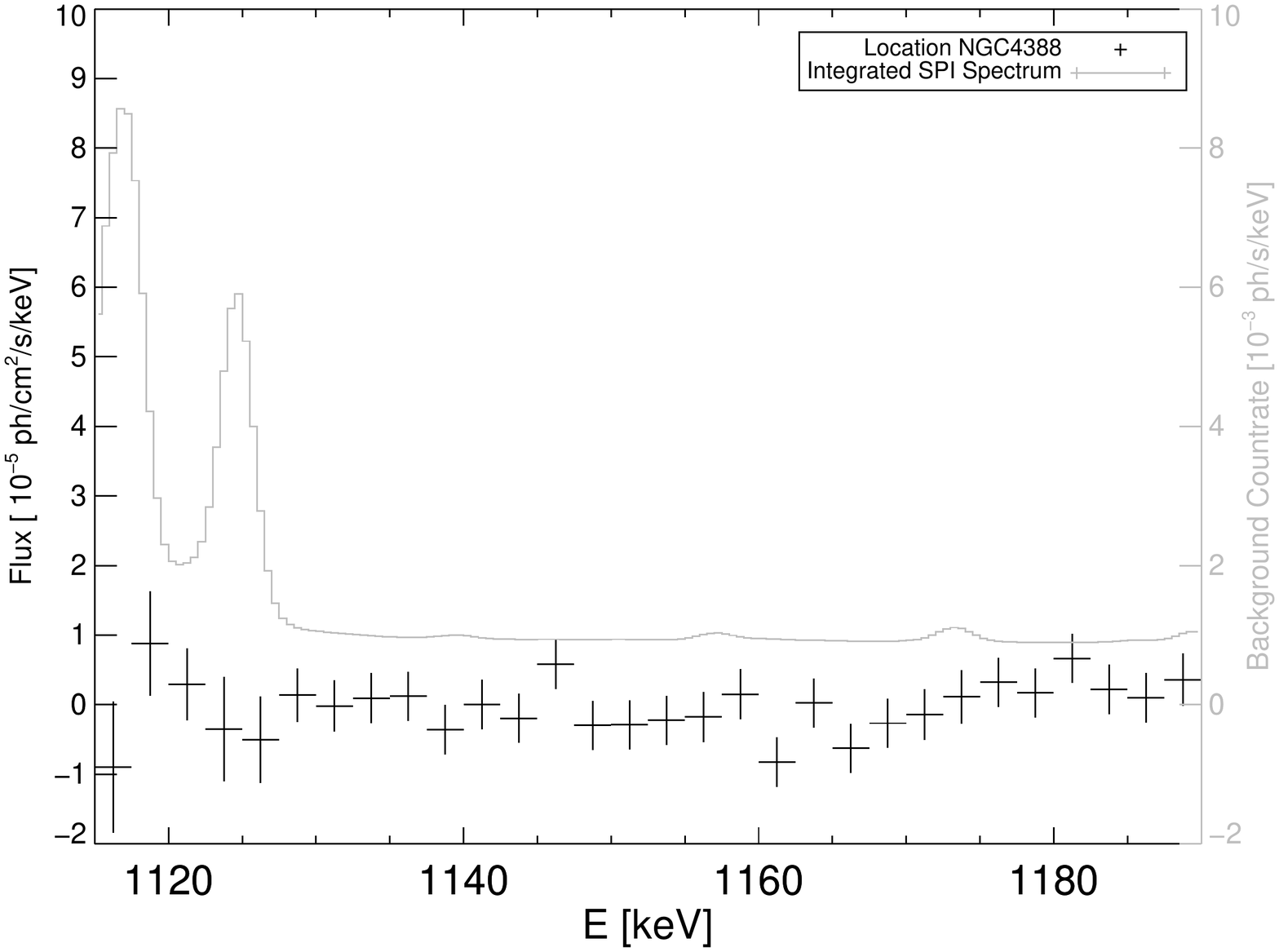} 
        \caption{Spectrum of galaxy NGC 4388 in black. The gray spectrum shows the average background count rate. Clearly visible are the strong background lines between 50 and 68\,keV and around 92\,keV. Our background modeling approach efficiently suppresses contribution from strong lines.}
        \label{fig:NGC4388}
\end{figure}
\begin{figure}
         \includegraphics[width=\linewidth,trim={1.5cm 1.5cm 1.0cm 1.5cm}]{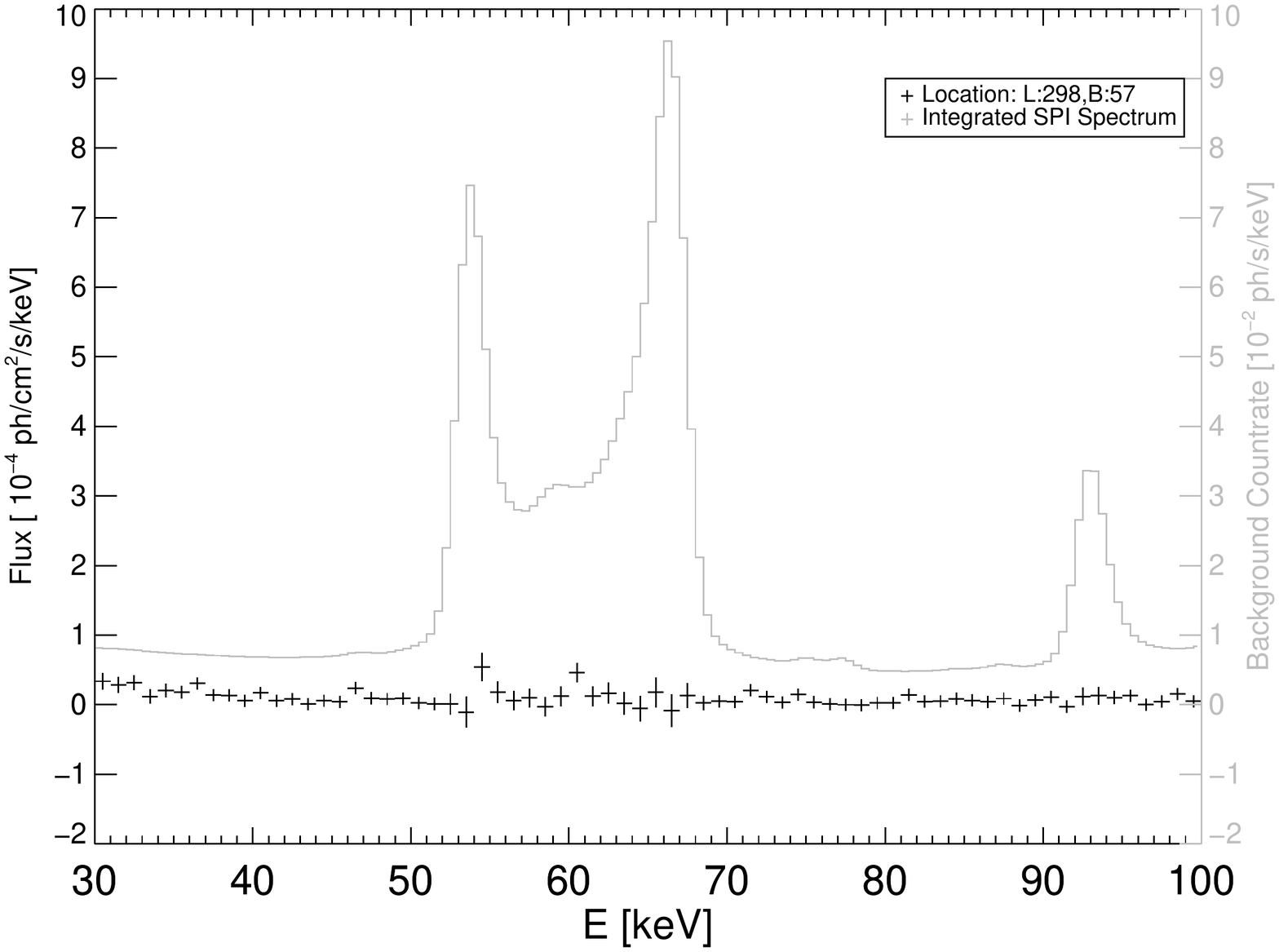}
          \caption{Spectra of an arbitrarily chosen celestial location with no known sources present. The gray line shows the average count rate for the background. Influence from strong background lines is suppressed by our modeling approach. Larger errors on the fit results are found in accordance with statistical analysis for energy bins with strong background. Dead time corrected exposure at the positions is 9\,Ms.}
         \label{fig:Sources_LowE}
\end{figure}
Figures \ref{fig:NGC4388} and \ref{fig:Sources_LowE} show the spectra for the galaxy cluster NGC4388 and "empty" space. With our background modeling approach, the strong background lines in the region between 50 and 68\,keV, 90\,keV, as well as 1115-1125\,keV are adequately suppressed. The photon index of the continuum emission ($\gamma = 1.740\pm0.074$) of NGC4388 (l = 279.1$^\circ$; b = 74.3$^\circ$) is consistent in the spectral index with the measurements of $1.72 \pm 0.05$ \citep{Beckmann2004}. We found no spurious $^{44}\rm Ti$ signatures in either location. In both regions, fluctuations around the expected continuum and baseline are present, however, these fluctuations are on scales smaller than the instrumental energy resolution at the respective energies and compatible with statistical fluctuations.

\end{document}